\documentclass[prd,showpacs,floatfix,amsmath,nofootinbib,amssymb,floatfix,twocolumn]{revtex4} 
\usepackage{graphicx,color,dcolumn,booktabs,bm}
\usepackage{longtable,lscape}
\usepackage{subfigure}
\usepackage{txfonts}
\usepackage{overpic}
\usepackage{amssymb,amsmath}
\usepackage{indentfirst}
\usepackage{feynmf}   
\usepackage{slashed}  
\usepackage{cases}
\usepackage{color}
\usepackage{multirow}
\usepackage{epstopdf}
\usepackage{longtable}
\usepackage{graphicx,color,dcolumn,booktabs,bm}
\usepackage[colorlinks,
            citecolor=blue,
            anchorcolor=red,
            menucolor=red,
            linkcolor=red,
            filecolor=red,
            runcolor=red,
            urlcolor=blue,
            frenchlinks=red]{hyperref}
\usepackage{times}
\usepackage{braket}
\usepackage{amsfonts,amssymb,stmaryrd,latexsym,amsmath}

\newcommand{\Tr}{\text{Tr}}

\def\DD{{DD^{*}}}
\def\DDbar{{D\bar{D}^{*}/\bar{D}D^*}}

\graphicspath{{Figures/}} %

\allowdisplaybreaks

\begin{document}

\title{Study of exotic hadron states in the $DD^{*}$ system via the complex momentum representation and Green's function method}

\author{Di Wu$^{1}$}
\author{Mao Song$^{1}$}
\email{songmao@ahu.edu.cn}
\author{Jian-You Guo$^{1}$}
\author{Gang Li$^{1}$}
\author{Xuan Luo$^{1}$}
\author{Peng Wang$^{1}$}

\affiliation{$^1$School of Physics, Anhui University, Hefei 230601, China}

\begin{abstract}

In this paper, we propose a novel approach to investigate exotic hadronic states. For the $DD^{*}$ system, we employ the projection operator method to derive the momentum-space interaction potential. Subsequently, the complex momentum representation (CMR) method is adopted to realize a unified description of bound states, resonant states, and the continuum. By combining the Green's function and the CMR, the scattering phase shifts and cross sections are determined. This integrated approach provides a comprehensive framework for analyzing the scattering dynamics of the $DD^{*}$ system. In the hadronic molecular state framework, the $X(3872)$, $T_{cc}^+$, and $Z_c(3900)$ states can be consistently explained as bound states, while the $G(3900)$ can be interpreted as a $P$-wave resonant state. The decomposition of the scattering phase shifts and cross sections facilitates understanding the roles of resonant and continuum spectrum.
\end{abstract}


\maketitle

\section{introduction}\label{sec1}
The study of exotic hadrons is a prominent frontier in particle physics. These hadrons fundamentally challenge the conventional quark model while deepening our understanding of Quantum Chromodynamics (QCD). While hadrons were long classified as simple baryons ($qqq$) and mesons ($q\bar{q}$), QCD inherently permits more complex structures known as exotic states~\cite{Gell-Mann:1964}. The discovery of the $X(3872)$~\cite{Belle:2003nnu,CDF:2003cab,D0:2004zmu,BaBar:2004oro} in 2003 by the Belle Collaboration revealed a state with a mass near the $D^0\bar{D}^{\ast 0}$ threshold and incompatible with charmonium, thereby igniting widespread interest in exotic hadrons. Subsequent experiments, including BaBar and LHCb, have since identified unambiguous exotic states such as $X$, $Y$, $Z$, and $P_c$ states, firmly establishing physics beyond the quark model. The nature of these exotic states remains debated, and exploring their structures and interactions remains a key focus in hadron physics.

\par
Various phenomenological models, such as the chiral effective field theory~\cite{Xu:2017tsr,Ohkoda:2012hv,Li:2012cs,Ren:2021dsi}, Bethe-Salpeter approach~\cite{Sakai:2017avl,He:2014nya,He:2015mja,Wallbott:2019dng}, constituent quark models~\cite{Zhu:2019iwm,Ortega:2021yis,Tan:2020ldi,Luo:2017eub}, QCD sum rules~\cite{Navarra:2007yw,Xin:2021wcr,Tang:2019nwv}, and relativized quark models~\cite{Lu:2020rog,Ebert:2007rn,Wang:2018pwi}, have been proposed to describe these exotic states.
In these models, the hadronic molecule and tetraquark interpretations provide clear physical insights into the internal structure of exotic states. Especially in the vicinity of the threshold, the hadronic molecule picture is often considered one of the candidates for the interpretation of these structures.

Until now, three exotic hadron states $X(3872)$, $T_{cc}^+$ and $Z_c(3900)$ have been discovered experimentally, with masses near the $DD^*$ threshold.
It is well known that the $X(3872)$ is widely interpreted as a loose bound state of $D\bar{D}^{*}$ with quantum number $J^{PC} = 1^{++}$ ~\cite{Wang:2013kva,Thomas:2008ja,Braaten:2010mg}. The ratio $\mathcal{B}[X(3872)\to J/\psi \pi^+\pi^-\pi^0]/\mathcal{B}[X(3872)\to J/\psi \pi^+\pi^-]$ indicates significant isospin breaking in the hidden-charm decay of $X(3872)$ \cite{D0:2004zmu,Belle:2005lfc,BaBar:2010wfc}. $T^+_{cc}(cc\bar{u}\bar{d})$ is considered as a molecular structure of $DD^{*}$ with quantum number $I(J^P) = 0(1^+)$~\cite{Sakai:2023syt,Ohkoda:2012hv,Cheng:2022qcm,Ren:2021dsi,Liu:2019stu}. $Z_c(3900)$ is a charged hidden-charm exotic hadron first observed by the BESIII collaboration in 2013, with a mass of 3887.1 MeV and a decay width of 28.4 MeV. Although its mass lies slightly above the $D\bar{D}^*$ threshold, several studies indicate that its pole position can shift by tens of MeV~\cite{Yu:2024k,Chen:2023def,Zhu:2024hgm,Nakamura:2023obk}. Despite the fact that the interaction of $I=1$ is weaker than that of $I=0$ in $D\bar{D}^*$ systems, $Z_c(3900)$ is also suggested to be an isovector $D\bar{D}^*$ molecule state with quantum numbers $J^{PC}=1^{+-}$~\cite{Guo:2013sya,Wang:2013cya}. Recent theoretical studies suggest that $G(3900)$ can be interpreted as a $P$-wave resonant state of the $D\bar{D}^*/\bar{D}D^*$~\cite{Lin:2024prl,Chen:2025gxe,Liu:2025sjz}. However, the scattering amplitudes global analysis for the processes $e^+e^- \to D\bar{D},D\bar{D}^*+c.c.$, and $D^*\bar{D}^*$ indicates the $G(3900)$ as a dynamically generated state~\cite{Ye:2025ywy}. The $D^{(*)}D^{(*)}$ and $D^{(*)}\bar{D}^{(*)}$ $S$-wave bound states have been extensively studied within the one-boson-exchange (OBE) model~\cite{Li:2012cs,Li:2012ss,Liu:2019stu,Abreu:2022sra,Abreu:2015jma}. However, $P$-wave resonant states remain relatively less explored and have attracted only limited research attention until recently~\cite{Lin:2024prl,Whyte:2024ihh}.

Resonance phenomena are pervasive in nature, appearing in atomic physics, molecular physics, nuclear physics, and chemical reactions. Therefore, a series of scattering theory methods have been proposed, such as R-matrix~\cite{Wigner:1947zz, Hale:1987zz}, K-matrix~\cite{Humblet:1991zz}, J-matrix~\cite{Taylor}, scattering phase shift analysis, continuous spectrum theory, and coupled-channel method. However, to address the complexities of these approaches, several bound-state-like methods have been developed, which include the real stabilization method~\cite{Hazi}, analytic continuation of the coupling constant~\cite{Kukulin}, and complex scaling method (CSM)~\cite{csm1,csm2}. The CSM provides a unified framework for describing bound states, resonant states, and continuum states, making it widely applicable to resonance studies in atomic, molecular, and nuclear physics. Based on its advantages, the CSM has been widely adopted in hadron physics~\cite{Yu:2021lmb,Wang:2022yes,Cheng:2022qcm,Wang:2023ivd,Lin:2024prl,Lin:2022wmj,Chen:2023eri,Mei:2022msh,Lu:2025zae}. However, the CSM still has certain limitations. In the CSM, multiple diagonalizations of the Hamiltonian are required to accurately identify resonance parameters. Furthermore, the numerical results are often sensitive to unphysical parameters, such as the complex rotation angle.

To resolve these drawbacks, the complex momentum representation (CMR) method~\cite{1968On,Deltuva2015Momentum,2006Study} offers a viable alternative.
This method not only enables a unified description of bound states, resonant states, and continuum states~\cite{Sukumar:1978rf,Kwan:1978zh,Li:2016gbp}, but also yields more stable and precise numerical results without introducing any unphysical parameters. In Ref.~\cite{He:2025svv}, we firstly extended the CMR to the investigation of exotic hadronic states.

Typically, the interaction potential between hadrons is derived in momentum space within the one-boson-exchange (OBE) approximation and then transformed into coordinate space to solve the Schr\"{o}dinger equation. In contrast, the CMR permits direct solution of the Schr\"{o}dinger equation in momentum space, avoiding the requirement for Fourier transforms. For potentials with simple spin structures, the partial-wave expansion is adopted in Ref.~\cite{He:2025svv} to separate the potential components of different angular momenta. However, for systems with complicated spin configurations, it is necessary to evaluate transition matrix elements among various coupled channels, following similar procedures as in coordinate space. Here, we propose a novel projection operator method to efficiently construct momentum-space potentials for systems with complex spin structures~\cite{Meng:2023bmz,Chacko:2024cax}. When combined with the Green's function method, the CMR framework can be extended to investigate scattering phase shifts and cross sections, providing a unified theoretical description of resonance phenomena~\cite{Zhang:2025pro,Odsuren:2014sna}.

In this work, the CMR combined with the Green's function method is employed to investigate the resonant states of $G(3900)$. Within this framework, scattering observables such as phase shifts can be decomposed into resonant and background contributions. This decomposition enables the extraction of pure resonance components, offering deeper physical insight into the internal structure of such states. As demonstrated by Suzuki et al.~\cite{Ryusuke}, scattering phase shifts can be derived from the continuum level density, which is constructed via the complex momentum Green's function. The continuum level density (CLD), also known as the time delay function, is directly related to the $S$-matrix~\cite{Shlomo}. In the CMR, the discretized continuum is described by complex-energy eigenstates. Using these complex eigenenergies, the CLD can be evaluated as a smooth function of the real energy axis. This integrated approach offers valuable insights into the study of resonance and scattering phenomena in exotic hadronic states.

This paper is organized as follows. After the introduction, Section \ref{sec2} provides the theoretical framework and calculation methods. Next, we present the frameworks for the continuum level density, scattering phase shifts, and cross sections in Section \ref{sec3}. Section \ref{sec4} presents the numerical results and discussion, followed by a summary in Section \ref{sec5}.

\section{Theoretical framework}\label{sec2}
Based on the heavy quark symmetry and chiral symmetry of Quantum Chromodynamics (QCD), the interactions between the heavy mesons $D$ and $D^*$ can be systematically described within the framework of Heavy Meson Chiral Perturbation Theory (HM$\chi$PT). Heavy quark spin symmetry (HQSS) implies that the pseudoscalar meson $P$ and vector meson $P^*$ are degenerate in the infinite mass limit, allowing them to be unified into a single field $\mathcal{H}$~\cite{Liu:2009qhy,Nambu:1961tp,Burdman:1992gh,Wise:1992hn,Yan:1992gz,Casalbuoni:1996pg,Manohar:2000dt,Isola:2003fh,Li:2012cs}, defined as:

\begin{equation}
        {\cal H}=\frac{1+\slashed{v}}{2}(P_{\mu}^*\gamma^{\mu}-P\gamma_{5}),
\end{equation}
where $P=(D^0,D^+)$ and $P_\mu^*=(D^{*0},D^{*+})_\mu$ denote the pseudoscalar and vector fields, respectively, and $v=(1,0,0,0)$ represents the four-velocity of the heavy meson. The corresponding antiparticle superfield, $\tilde{{\cal H}}$ is given as:
\begin{eqnarray}
  \tilde{{\cal H}}=(\tilde{P}_{\mu}^*\gamma^{\mu}-\tilde{P}\gamma_{5})\frac{1-\slashed{v}}{2},
\end{eqnarray}
where $\tilde{P}=(\bar{D}^0,D^-)^\text{T}$ and $\tilde{P}_\mu^*=(\bar{D}^{*0},D^{*-})_\mu^\text{T}$. Under the charge conjugation operation $\mathcal{C}$, we adopt the convention: $D\xrightarrow{\mathcal{C}}\bar{D}$ and $D^*\xrightarrow{\mathcal{C}}-\bar{D}^*$, such that the superfield transforms as $\mathcal{H}\xrightarrow{\mathcal{C}}C^{-1}\tilde{\mathcal{H}}^\text{T}C$, where $C=i\gamma^2\gamma^0$ is the charge conjugation matrix. The conjugations of $\cal{H}$ and $\tilde{\mathcal{H}}$ are defined as $\bar{\mathcal{H}}=\gamma_0\mathcal{H}^\dagger\gamma_0$ and $\bar{\tilde{\mathcal{H}}}=\gamma_0\tilde{\mathcal{H}}^\dagger\gamma_0$, respectively.

In the One-Boson-Exchange (OBE) model, the effective Lagrangians that describe the interactions between heavy mesons and light mesons are formulated as follows:
\begin{eqnarray}
    &\mathcal{L}&=g_s\Tr\left[\mathcal{H}\sigma\bar{\mathcal{H}}\right]+ig_a\Tr\left[\mathcal{H}\gamma_\mu\gamma_5\mathcal{A}^\mu\bar{\mathcal{H}}\right]\nonumber\\
    &&+i\beta\Tr\left[\mathcal{H} v_\mu (\mathcal{V}^\mu-\rho^\mu)\bar{\mathcal{H}}\right]+i\lambda\Tr\left[\mathcal{H}\sigma_{\mu\nu}F^{\mu\nu}\bar{\mathcal{H}}\right]\nonumber\\
    &&+g_s\Tr\left[\bar{\tilde{\mathcal{H}}}\sigma\tilde{\mathcal{H}}\right]+ig_a\Tr\left[\bar{\tilde{\mathcal{H}}}\gamma_\mu\gamma_5\mathcal{A}^\mu\tilde{\mathcal{H}}\right]\nonumber\\
    &&-i\beta\Tr\left[\bar{\tilde{\mathcal{H}}} v_\mu (\mathcal{V}^\mu-\rho^\mu)\tilde{\mathcal{H}}\right]+i\lambda\Tr\left[\bar{\tilde{\mathcal{H}}}\sigma_{\mu\nu}F^{\mu\nu}\tilde{\mathcal{H}}\right].
\end{eqnarray}
where $F^{\mu\nu}=\partial^\mu\rho^\nu-\partial^\nu\rho^\mu-[\rho^\mu,\rho^\nu]$ denotes the field strength tensor of vector mesons.
The vector and axial-vector currents of the pseudoscalar Goldstone bosons, $\mathcal{V}^\mu$ and $\mathcal{A}^\mu$, are defined as:
\begin{eqnarray}
    &&\mathcal{V}^\mu=\frac{1}{2}[\xi^\dagger,\partial_\mu\xi],\; \mathcal{A}^\mu=\frac{1}{2}\{\xi^\dagger,\partial_\mu\xi\},\;
    \xi=\exp(i\mathbb{P}/f_\pi).
\end{eqnarray}
Similarly, the multiplet for the vector meson fields $\mathbb{P}$ is given by:
\begin{eqnarray}
\mathbb{P}=
\begin{bmatrix}\frac{\pi^{0}}{\sqrt{2}}+\frac{\eta}{\sqrt{6}} & \pi^{+}\\
\pi^{-} & -\frac{\pi^{0}}{\sqrt{2}}+\frac{\eta}{\sqrt{6}}
\end{bmatrix}.\nonumber\\
\end{eqnarray}
The multiplet of the vector meson fields $\rho^\mu$ is
\begin{eqnarray}
     \rho^\mu=\frac{ig_V}{\sqrt{2}}\begin{bmatrix}\frac{\rho^{0}+\omega}{\sqrt{2}}
 & \rho^{+}\\
\rho^{-} & \frac{-\rho^{0}+\omega}{\sqrt{2}}
\end{bmatrix}^\mu.
\end{eqnarray}

While the isospin triplets ($\rho, \pi$) and isospin singlets ($\omega, \eta$) belong to different representations under the flavor $SU(2)$ symmetry, we adopt the standard approach of embedding them into unified matrices. This treatment accounts for their underlying flavor $SU(3)$ relationships and helps reduce the number of independent coupling constants in the model.

To derive the $\mathrm{P}^{(\ast)}\mathrm{P}^{(\ast)}$ potential, effective Lagrangians are introduced to describe the interactions between heavy mesons mediated by the exchange of the pseudoscalar meson $\pi$, vector mesons ($v =\rho ,\omega$) and scalar meson $\sigma$~\cite{Ding:2008gr,Sakai:2023syt,Ohkoda:2012hv}. The effective potentials for the $DD^*$ system in momentum space are listed as follows~\cite{Lin:2024prl,Zhu:2024hgm}:
\begin{eqnarray}
    &V_{\sigma}^{D}(\vec{p}',\vec{p})&=-\frac{g_{s}^{2}\vec{\epsilon}\cdot\vec{\epsilon}'^{*}}{\vec{q}^{2}+m_{\sigma}^{2}},\nonumber\\
    &V_{\rho/\omega}^{D}(\vec{p}',\vec{p})&=\frac{\frac{1}{4}\beta^{2}g_{V}^{2}(\vec{\epsilon}\cdot\vec{\epsilon}'^{*})}{\vec{q}^{2}+m_{\rho/\omega}^{2}}\times\begin{cases}
        \tau\cdot\tau,\quad \text{for }\rho,\\
        \bm{1}\cdot\bm{1},\quad \text{for }\omega,
    \end{cases}\nonumber\\
    &V_{\pi}^{C}(\vec{p}',\vec{p})&=-\frac{g^{2}}{2f_{\pi}^{2}}\frac{(\vec{\epsilon}\cdot \vec{k})(\vec{\epsilon}'^{*}\cdot \vec{k})}{\vec{k}^{2}-k_0^2+m_{\pi}^{2}}\tau\cdot\tau,\nonumber\\
    &V_{\eta}^{C}(\vec{p}',\vec{p})&=-\frac{g^{2}}{6f_{\pi}^{2}}\frac{(\vec{\epsilon}\cdot \vec{k})(\vec{\epsilon}'^{*}\cdot \vec{k})}{\vec{k}^{2}-k_0^2+m_{\eta}^{2}}\bm{1}\cdot\bm{1},\nonumber\\
    &V_{\rho/\omega}^{C}(\vec{p}',\vec{p})&=\frac{\lambda^{2}g_{V}^{2}}{\vec{k}^{2}-k_0^2+m_{\rho/\omega}^{2}}\{(\vec{k}\cdot\vec{\epsilon})(\vec{k}\cdot\vec{\epsilon}'^{*})\nonumber\\
    &&-\vec{k}^{2}(\vec{\epsilon}\cdot\vec{\epsilon}'^{*})\}\times\begin{cases}
        \tau\cdot\tau,\quad \text{for }\rho,\\
        \bm{1}\cdot\bm{1},\quad \text{for }\omega,\label{potential}
    \end{cases}
\end{eqnarray}
where $D$ and $C$ denotes the direct and cross diagrams, respectively. The momentum-space potential can be transformed into coordinate space by Fourier transformation and decomposed into the central potential $C(r)$, tensor potential $T(r)$, and short-range contact potential $\delta(r)$, as shown in the Appendix~\ref{appendix;H}. In relevant previous studies, some theoretical calculations retain the short-range contact term, while others neglect it entirely. The CMR method allows the short-range term $\delta$ to be handled conveniently and self-consistently directly in momentum space. In the results and discussion section, we perform calculations for two cases, i.e., with and without the short-range contact term $\delta$, and compare the corresponding results in detail.

A form factor is introduced to regularize the meson-exchange potential by suppressing high-momentum (short-range) contributions. This reflects the physical reality that light mesons interact with heavy mesons as composite systems with finite spatial extension, rather than as point-like particles probing their internal structure. In this work, we implement a monopole form factor, $\mathcal{F}(\vec{q}, m)$, at each interaction vertex, which is defined as:
\begin{eqnarray}
\mathcal{F}(\vec{q};m)\!=\! \frac{\Lambda^{2} \!-\!
m^{2}}{\! \vec{q}^{\,\,2}+\Lambda^{2} \!}  \, ,
\end{eqnarray}
Here, $\Lambda$ denotes the cutoff parameter, while $m$ and $\vec{q}$ represent the mass and the three-momentum of the exchanged meson ($= \pi, \rho, \omega, \sigma$), respectively. Following the Refs.~\cite{Yasui:2009bz, Yamaguchi:2011xb, Ohkoda:2011vj}, $\Lambda$ is physically correlated with the root-mean-square (RMS) radius of the source hadron.

The Schr\"{o}dinger equation in momentum space can be expressed as follows,

\begin{equation}
\frac{\hbar^2k^2}{2\mu}\psi(\vec{k})+\int \frac{1}{(2\pi)^{3}} V\left(k,k'\right)\psi(\vec{k}')d\vec{k}'=E\psi(\vec{k}).\label{three-Schrodinger2}
\end{equation}%
Here, $\psi(\vec{k})$ represents the momentum wavefunction, with wavevector $\vec{k}$ corresponding to momentum.  The Schr\"{o}dinger equation has become an integral equation in momentum space.

Instead of directly solving the three-dimensional integral equation given in Eq.~(\ref{three-Schrodinger2}), we decompose it into radial and angular parts by invoking a partial wave expansion. The wavefunction $\psi(\vec{k})$ can be expanded in a complete set of spherical harmonics as
\begin{eqnarray}
\psi\left(\vec{k}\right)=f^l\left(k\right)Y_{lm}\left(\Omega_k\right).
\end{eqnarray}

The potentials in momentum space corresponding to different quantum numbers can be derived using the projection operator method. The individual partial waves are labeled as $^{2S+1}L_J$ with $S$, $L$ and $J$ denoting the total spin, the orbital angular momentum and the total momentum, respectively. The partial wave projection of the potentials is performed following the formalism established in Ref.~\cite{Chacko:2024cax}, which gives
\begin{eqnarray}
V_{\alpha\beta}(J^{PC}) = \frac{1}{2J + 1} \int \frac{d\Omega_n}{4\pi} \frac{d\Omega_{n'}}{4\pi} \mathrm{Tr}\left[ P^\dagger(\alpha, \vec{n}) V P(\beta, \vec{n}') \right] ,
\end{eqnarray}
where $\vec{n} = \vec{k}/|\vec{k}|$, $\vec{n}' = \vec{k}'/|\vec{k}'|$, $P^\dagger(\alpha, \vec{n})$ and $P(\beta, \vec{n}')$ are outgoing and incoming normalised projectors respectively with $\alpha$ and $\beta$ being the bases states. The projectors are normalised as:

\begin{eqnarray}
\int \frac{d\Omega_n}{4\pi} P^\dagger(\alpha, \vec{n}) P(\alpha, \vec{n}) = 2J + 1 \ .
\end{eqnarray}

Due to the spatial symmetry of the $2 \to 2$ scattering process, the potential depends only on the scattering angle $\theta$ between the initial and final relative momenta. By defining $x = \hat{k} \cdot \hat{k}' = \cos\theta$, the expression simplifies to:

\begin{eqnarray}
V_{\alpha\beta}\left(J^{PC}\right)=\frac{1}{2J+1}\int_{-1}^{+1}\frac{dx}{2}\mathrm{Tr}\left[P^\dagger(\alpha,\vec{n}') V P(\beta,\vec{n})\right] ,\label{project}
\end{eqnarray}
where the trace is performed over the indices of angular momentum, since $J$ is conserved. The relevant projectors are given by:
\begin{eqnarray}
    P[^3S_1]_{i} &=& \epsilon_i, \\
    P[^3D_1]_{i} &=& -\frac{3}{\sqrt{2}} \epsilon_i \left( n_i n_j - \frac{1}{3}\delta_{ij} \right), \\
    P[^3P_1]_{i} &=& \frac{\sqrt{6}}{2} i\epsilon_{ijk}\epsilon^*_j n_k.
\end{eqnarray}

After integrating over the solid angles, the Schr\"{o}dinger equation in momentum space becomes
\begin{equation}
\frac{\hbar^2k^2}{2\mu}f_l^J(k)+\int_0^\infty \frac{4\pi}{(2\pi)^{3}} V_{ll'}\left(k,k'\right)f_{l'}^J(k')k'^2dk'=Ef_l^J(k).\label{Schrodinger}
\end{equation}%

This possesses numerous advantageous characteristics. Firstly, it explicitly provides the most realistic hadron-hadron interactions derived from effective field theory in momentum space. Secondly, the boundary conditions imposed on the differential equation in coordinate space are seamlessly incorporated into the integral equation. Lastly, but importantly, integral equations are straightforward to implement numerically, and convergence can be achieved simply by increasing the number of integration points.

In order to solve the Schr\"{o}dinger equation numerically, the Gaussian-Legendre quadrature method is used to handle the momentum space integral.
Since the Hamiltonian matrix is asymmetrical, we make it symmetrical by the transformation
\begin{equation}
  \mathbf{f}(k_a)=\sqrt{w_a}k_a f(k_a)
\end{equation}
The Schr\"{o}dinger equation becomes
\begin{equation}
\frac{\hbar^2k_a^2}{2\mu}\mathbf{f}_l^J(k_a)+\sum_b \frac{4\pi k_ak_b\sqrt{w_aw_b}}{(2\pi)^{3}} V_{ll'}(k_a,k_b)\mathbf{f}_{l'}^J(k_b)=E \mathbf{f}_l^J(k_a).
\end{equation}

Then, the density in the momentum space is,
\begin{equation}
 \rho^J(k_a)=\sum_l \mathbf{f}^{*J}_l(k_a)\mathbf{f}^J_l(k_a).
\end{equation}

\section{Continuum level density and scattering phase shift}\label{sec3}
The level density $\rho(E)$ of the full Hamiltonian $H$ is defined as:
\begin{eqnarray}
\rho(E)&=&\sum_i\int\delta(E-E_i),\label{density}
\end{eqnarray}
where $E_i$ denotes the eigenvalues of the Hamiltonian $H$. Here, the summation and integration represent the contributions from the discrete and continuous spectra, respectively. Using the Green's function approach, the level density of the full Hamiltonian $H$ can be equivalently expressed as:
\begin{eqnarray}
\rho(E)=-\frac{1}{\pi}\text{Im}\left\{\text{Tr}\left[\frac{1}{E+\text{i}0-H}\right]\right\}. \label{density2}
\end{eqnarray}

When the Hamiltonian $H$ is described as the sum of an asymptotic term $H_0$ and the interaction term $V$, namely $H=H_0+V$, the continuum level density $\Delta(E)$ can be determined by the density $\rho(E)$ obtained from the full Hamiltonian $H$ and level density $\rho_0(E)$ of continuum states obtained from the asymptotic Hamiltonian $H_0$ as follows~\cite{shlomo1992, Levine:1969, Odsuren:2014,Zhang:2025}:

\begin{eqnarray}
\Delta(E)&=&\rho(E)-\rho_0(E) \\ \nonumber
&=& -\frac{1}{\pi}\text{Im}\left[\text{Tr}\left\{\frac{1}{E+\text{i}0-H}-\frac{1}{E+\text{i}0-H_0}\right\}\right]. \label{density3}
\end{eqnarray}

The connection between the continuum level density (CLD) and the scattering $S$-matrix is established via the Krein-Birman formula:
\begin{eqnarray}
\Delta(E) &=& \frac{1}{2\pi}\text{Im}\frac{\mathrm{d}}{\mathrm{d}E}\ln\{\det S(E)\}, \label{delta-density}
\end{eqnarray}

For a single-channel system, the scattering $S$-matrix is typically parameterized as $S(E) = e^{2i\delta(E)}$, where $\delta(E)$ denotes the scattering phase shift. Under this representation, the CLD and its corresponding phase shift are related through the following simplified expressions~\cite{shlomo1992, Levine:1969, Odsuren:2014,Zhang:2025}:
\begin{eqnarray}
\Delta(E) &=& \frac{1}{\pi}\frac{\mathrm{d}\delta}{\mathrm{d}E}, \quad \delta(E) = \pi\int_{-\infty}^{E}\Delta(E')\mathrm{d}E'. \label{phase-shift}
\end{eqnarray}
This derivative relationship implies that the CLD effectively captures the rate of change in the phase shift. Thus, any resonant behavior characterized by a sharp increase in $\delta(E)$ will produces a corresponding peak in the spectral density.

In the CMR approach, the completeness relation (Berggren basis) is utilized to discretize the trace. The integration contour in the CMR is deformed to expose the resonance poles. By summing over the discretized eigenstates obtained via $N$ momentum mesh points, the CMR-CLD at a given energy $E$ is formulated as:

\begin{eqnarray}
\Delta_N^{\text{CMR}}(E) &=& \sum_{b}^{N_b}\delta(E - E_b) + \frac{1}{\pi}\sum_{r}^{N_r^{\text{CMR}}}\frac{\Gamma_r/2}{(E - E_r)^2 + \Gamma_r^2/4} \nonumber\\
&+& \frac{1}{\pi}\sum_{c}^{N_c^{\text{CMR}}=N-N_b-N_r^{\text{CMR}}}\frac{E_c^i}{(E - E_c^r)^2 + (E_c^i)^2} \nonumber\\
&-& \frac{1}{\pi}\sum_{c}^{N}\frac{E_c^{0i}}{(E - E_c^{0r})^2 + (E_c^{0i})^2}, \label{CMR-CLD}
\end{eqnarray}
where $E_b$, $E_r -\mathrm{i}\Gamma_r/2$ and $E_c^r -\mathrm{i}E^i$ are eigenvalues of the full Hamiltonian $H = T + V$ and $E_c^{0r} -\mathrm{i}E^{0i}$ are eigenvalues of asymptotic Hamiltonian $H_0 = T$. $N_b$ denotes the number of the bound states and $N_r^{\text{CMR}}$ represents the number of the resonant states.

According to Eq.~\ref{phase-shift}, the scattering phase shift, evaluated over the $N$ complex momentum mesh points in the CMR, is expressed as:
\begin{eqnarray}
\delta_N^{\text{CMR}}(E) &=& \pi \int_{-\infty}^{E} \Delta_N^{\text{CMR}}(E') \mathrm{d}E' \nonumber\\
&=& N_b \pi + \int_{-\infty}^{E} \mathrm{d}E' \sum_{r=1}^{N_r^{\text{CMR}}} \frac{\Gamma_r/2}{(E' - E_r)^2 + \Gamma_r^2/4} \nonumber\\
& +& \int_{-\infty}^{E} \mathrm{d}E'  \sum_{c=1}^{N_c^{\text{CMR}}} \frac{E_c^i}{(E' - E_c^r)^2 + (E_c^i)^2} \nonumber\\
&-& \int_{-\infty}^{E} \mathrm{d}E' \sum_{k=1}^{N} \frac{E_c^{0i}}{(E' - E_c^{0r})^2 + (E_c^{0i})^2} \nonumber\\
&=& N_b \pi +\sum_{r=1}^{N_r} \delta_r+\sum_{c=1}^{N_c} \delta_c-\sum_{k=1}^{N} \delta_k,\label{phase-shift2}
\end{eqnarray}
where $\delta_{r}$ represents the resonance phase shift, $\delta_{c}$ is the continuum phase shift, and $\delta_{k}$ denotes the background phase shift.
By integrating each term, we obtain the spectral decomposition of the phase shift:
\begin{eqnarray}
\cot\delta_r = \frac{E_r^{\text{res}} - E}{\Gamma_r/2},
\cot\delta_c = \frac{E_c^r - E}{\epsilon_E^i},
\cot\delta_k^0 = \frac{E_k^{0r} - E}{E_k^{0i}}.
\end{eqnarray}

The cross section can be described by using these phase shifts, and we can see that these contributions originate from bound states, resonant poles, and the continuum spectrum.
When focusing on the contribution of a single resonant pole, the remaining components may be collectively treated as a background phase shift. This framework is consistent with the analysis framework established by Fano~\cite{U.Fano}, as the partial cross section $\sigma_l(E)$ (corresponding to the orbital angular momentum $l$) is defined as

\begin{eqnarray}
\sigma_l(E) = \frac{4\pi(2l + 1)}{k^2} \sin^2\delta_l(E), \label{cross-section}
\end{eqnarray}
where $k^2 = 2E\mu$ and $\mu$ denotes the reduced mass of the scattering system. The phase shift $\delta_l(E)$ can be decomposed into two parts: $\delta_l(E) = \delta_r + \delta_B$, where $\delta_r$ denotes the contribution from the single resonance, and $\delta_B$ corresponds to the background term. The spectral shape of the cross section can thus be systematically analyzed by separately evaluating the resonance phase shift $\delta_r$ and the background phase shift $\delta_B$.

\section{Numerical results}\label{sec4}

In this section, we present our numerical results and provide a detailed analysis of several key physical quantities, including the probability distribution, energy density, scattering phase shift, and scattering cross section. The relevant parameters and meson masses are listed in Table~\ref{parameter}. Based on Ref.~\cite{Sakai:2023syt}, the $\sigma$-meson coupling constant $g_s$  is set to be one-third of the nucleon-$\sigma$ meson coupling strength. To validate our approach, we compare our results with those obtained via the complex scaling method in Ref.~\cite{Lu:2025zae}, and find good agreement between the two methods.
	
\begin{table}[!htbp]
	\caption{The relevant parameters are used in this work \cite{Sakai:2023syt}.}\label{parameter}
	\begin{tabular}{ccc|cc}\toprule[2pt]
		Hadron       &$I(J^P)$     &Mass (MeV)    &Parameters   & \\\hline
		$\pi$        &$1(0^-)$                    &138          &g   &  0.59  \\
		$\rho$       &$1(1^-)$          &770          &$g_V$     & $\frac{m_\rho}{\sqrt{2}f_\pi}$   \\
		$\omega$     &$0(1^-)$          &782          &$\beta$   & 0.9   \\
		$\sigma$     &$0(0^+)$          &500          &$\lambda$ & 0.56 $\rm{GeV^{-1}}$     \\
		$\mathrm{D}$        &$\frac{1}{2}(0^-)$   &1868         &$g_s$     & 3.4    \\
		$\mathrm{D}^{\ast}$  &$\frac{1}{2}(1^-)$  &2009         &$f_{\pi}$  &  93 MeV   \\
		\bottomrule[2pt]
	\end{tabular}
\end{table}

\begin{table}[htbp]
  \centering
  \caption{Cutoff parameter values at specific binding energies for the hadronic molecular states $X(3872)$, $T_{cc}(3875)$ and $Z_c(3900)$, with and without the short-range contact terms ($\delta$)}
	 	\begin{tabular}{lccc}
	 		\toprule[2pt] 
	 		  & $X(3872)$ & $T_{cc}(3875)$ & $Z_c(3900)$ \\\hline
	 		\midrule
	 		$I(J^{PC})$   & $0^+(1^{++})$ & $0(1^+)$ & $1^+(1^{+-})$ \\
	 		$E$ (MeV)    & $-5.031$& $-0.273$ & $-0.085$ \\
	 		($\delta$) $\Lambda$ (GeV)  & $0.8272$ & $0.749$ & $0.9998$ \\
	 		(no $\delta$) $\Lambda$ (GeV)  & $1.00527$ & $1.0348$ & $1.1940$ \\
	 		\bottomrule[2pt] 
	 	\end{tabular}
	 	\label{tab:cutoff_params}
\end{table}

For the description of hadronic molecular states, the form factor is the dominant source of theoretical uncertainty. We first adjust the cutoff parameter to reproduce the binding energies: $-5.031$ MeV for $X(3872)$, $-0.273$ MeV for $T_{cc}(3875)$, and $-0.085$ MeV for $Z_c(3900)$, and obtain the corresponding cutoff parameters both with and without the contact term $\delta$, as presented in Table \ref{tab:cutoff_params}. It can be observed that for the interaction potential with the contact term $\delta$, the cutoff parameters are $0.8272$, $0.749$ and $0.9998$ MeV for $X(3872)$, $T_{cc}(3875)$ and $Z_c(3900)$, respectively. When the contact term is not included, the corresponding cutoff parameters are $1.00527$, $1.0348$ and $1.1940$ MeV, respectively. By comparison, the cutoff parameters without the contact term $\delta$ are all around 1.0 GeV with smaller variations, which appears better suited to describe hadronic molecular states.

\begin{figure*}
	 	\centering
	 	\includegraphics[width=0.85\textwidth]{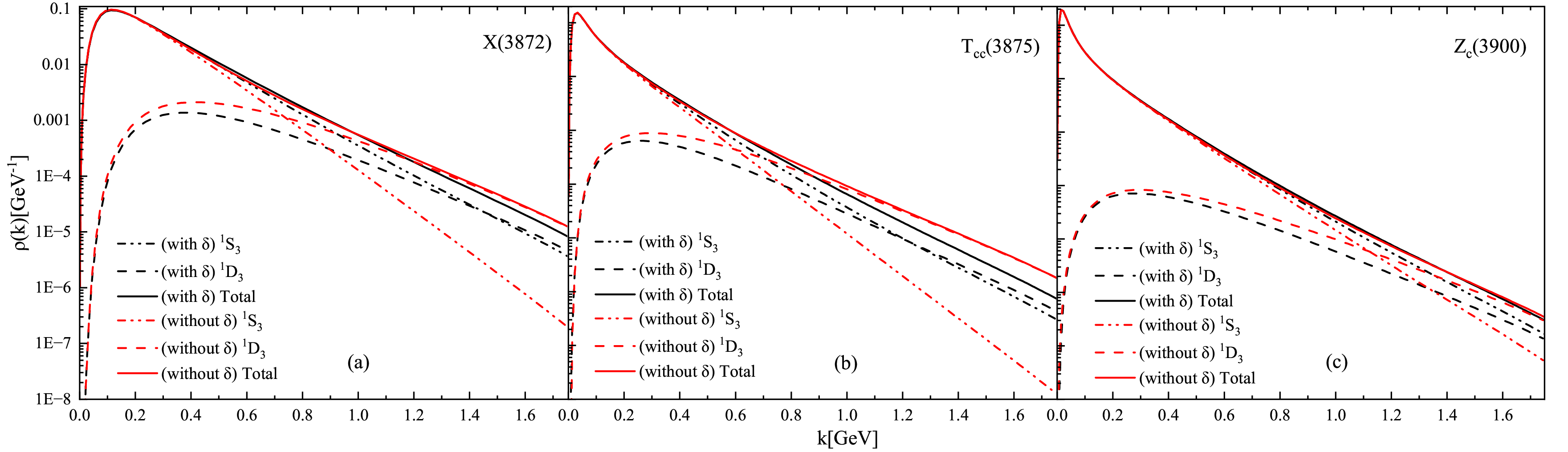}
	 	\caption{ (Color online) Density distribution of $X(3872)$, $T_{cc}(3875)$, and $Z_c(3900)$ with the $S\text{--}D$ mixing. The black and red curves correspond to the cases with and without the short-range contact term $\delta$, respectively. The double-dashed, dashed, and solid lines represent the $S$-wave, $D$-wave, and total $S$+$D$ contributions, respectively.}
	 	\label{density}
\end{figure*}

To further investigate the role of the contact term $\delta$, we compare the momentum-space density distributions of $X(3872)$, $T_{cc}(3875)$,  and $Z_c(3900)$ with and without the contact term $\delta$, as shown in Fig.\ref{density}. Here, the binding energies and cutoff parameters are set as in Table~\ref{tab:cutoff_params}. In Fig.\ref{density}, the black curve denotes the results obtained with the inclusion of the contact term, whereas the red curve denotes those obtained without it. The double-dashed, dashed, and solid lines represent the $S$-wave component, the $D$-wave component, and the total contribution, respectively. The results show that, whether the short-range contact term $\delta$ is included or not, the $S$-wave contribution dominates, exceeding $99\%$ of the total, while the $D$-wave contributes less than $1\%$ and can be neglected. To highlight the influence of the $D$-wave, the vertical axis is plotted on a logarithmic scale. As illustrated in Fig.\ref{density}, the $D$-wave contribution increases as the relative momentum $k$ increases, whereas the $S$-wave contribution gradually decreases. In the region of large relative momentum, the $D$-wave component even exceeds the $S$-wave contribution. By comparing Figs.\ref{density}(a)-(c), we find that the $D$-wave contribution exceeds the $S$-wave contribution at a smaller relative momentum when short-range contact terms are included.
	
\begin{figure}
	 	\centering
	 	\includegraphics[width=0.85\textwidth]{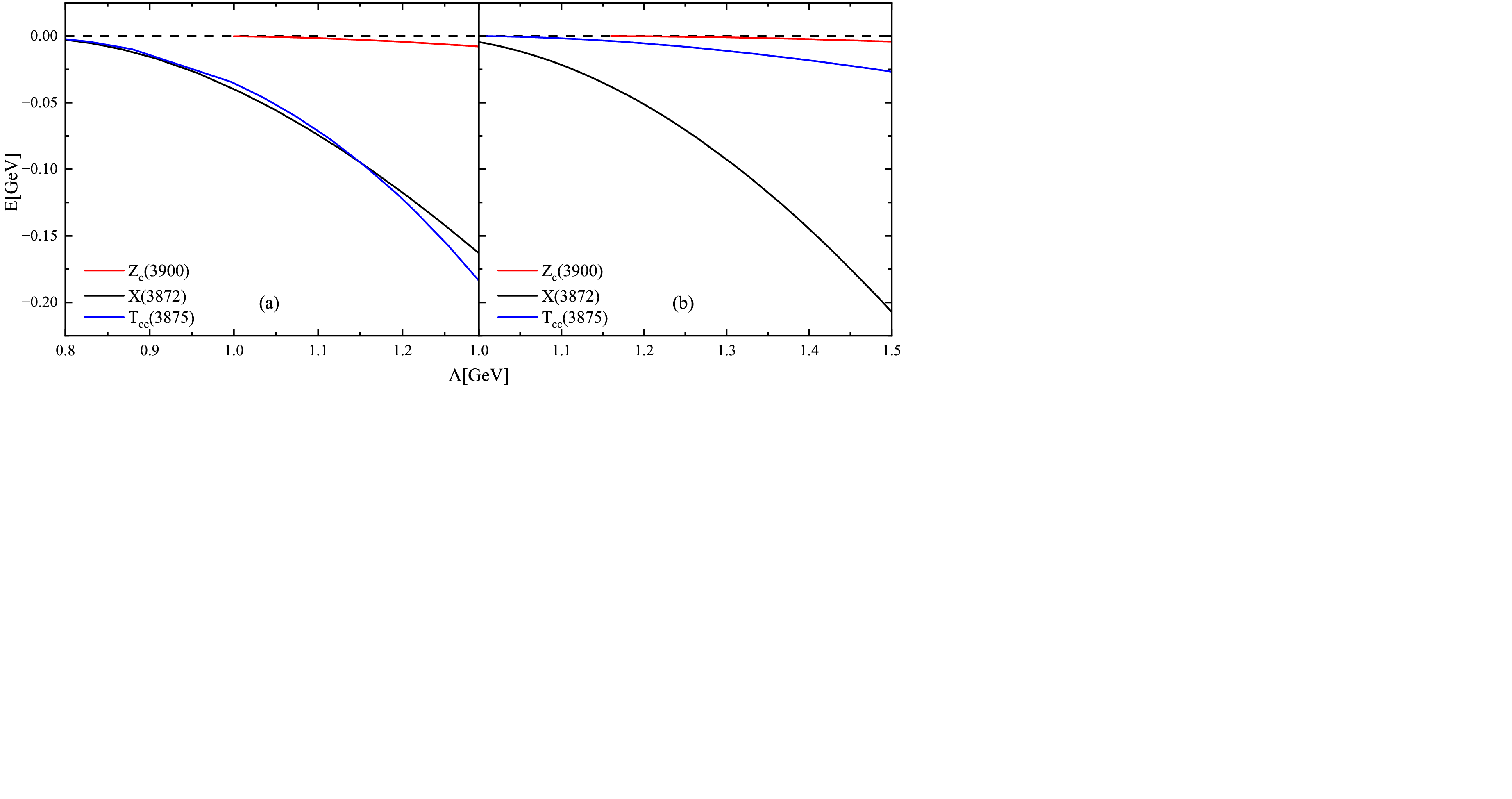}
         \vspace{-4.5cm}
	 	\caption{(Color online) The binding energies of $X(3872)$, $T_{cc}(3875)$, and $Z_c(3900)$ as functions of the cutoff parameter $\Lambda$. (a) with the short-range contact term $\delta$, (b) without the short-range contact term $\delta$.}
	 	\label{cutoff}
\end{figure}
	
The cutoff parameter of the monopole form factor is the dominant source of theoretical uncertainty. To more clearly illustrate the dependence on the cutoff parameter, we present the evolution of the binding energies of $X(3872)$, $T_{cc}(3875)$, and $Z_c(3900)$ as a function of the cutoff parameter $\Lambda$ in Fig.\ref{cutoff}(a) (with short-range contact term $\delta$) and Fig.\ref{cutoff}(b) (without short-range contact term $\delta$). The blue solid line, black solid line, and red solid line correspond to $T_{cc}(3875)$, $X(3872)$, and $Z_c(3900)$, respectively. Overall, the contact term $\delta$ provides an additional attractive interaction. In the range of cutoff parameters $\Lambda$, the $1^+(1^{+-})$ state forms a shallow bound state, which is consistent with the quantum numbers of the hadronic state $Z_c(3900)$. The inclusion of the contact term makes the system easier to form a bound state, and the binding energy is larger than that without the contact term $\delta$. Without the contact term, the binding energy decreases from 0 to -4.2 MeV as $\Lambda$ increases from 1.0 to 1.5 GeV. Meanwhile, with the contact term, the binding energy decreases from 0 to -8.24 MeV as $\Lambda$ increases from 0.8 to 1.3 GeV. The $0(1^+)$ state forms a deep bound state, corresponding to the hadronic state $X(3872)$. As the cutoff parameter $\Lambda$ increases, the binding energy increases significantly, and the contact term further enhances the binding energy. When the contact term is excluded, the binding energy of $X(3872)$ decreases from -4.45 to -207.2 MeV as $\Lambda$ increases from 1.0 to 1.5 GeV. However, with the contact term included, the binding energy decreases from -2.73 to -168.12 MeV as $\Lambda$ increases from 0.8 to 1.3 GeV. The contact term has a significant impact on the $0(1^+)$ state, which corresponding to the hadronic state $T_{cc}(3875)$. In the absence of the contact term $\delta$, a shallow bound state is formed, which exhibits only a weak dependence on the cutoff parameter $\Lambda$, the binding energy decreases from 0 to -26.7 MeV as $\Lambda$ increases from 1.0 to 1.5 GeV. When the contact term is included, the energy of the bound state exhibits a strong dependence on the cutoff parameter $\Lambda$, the binding energy of $T_{cc}(3875)$ decreases from -2.2 to -190.64 MeV as $\Lambda$ increases from 1.0 to 1.3 GeV. These results suggest that the $X(3872)$ and $T_{cc}(3875)$ states are more sensitive to the short-range contact term within the effective interaction.	

\begin{figure}
	\centering
	\includegraphics[width=3.33in, keepaspectratio]{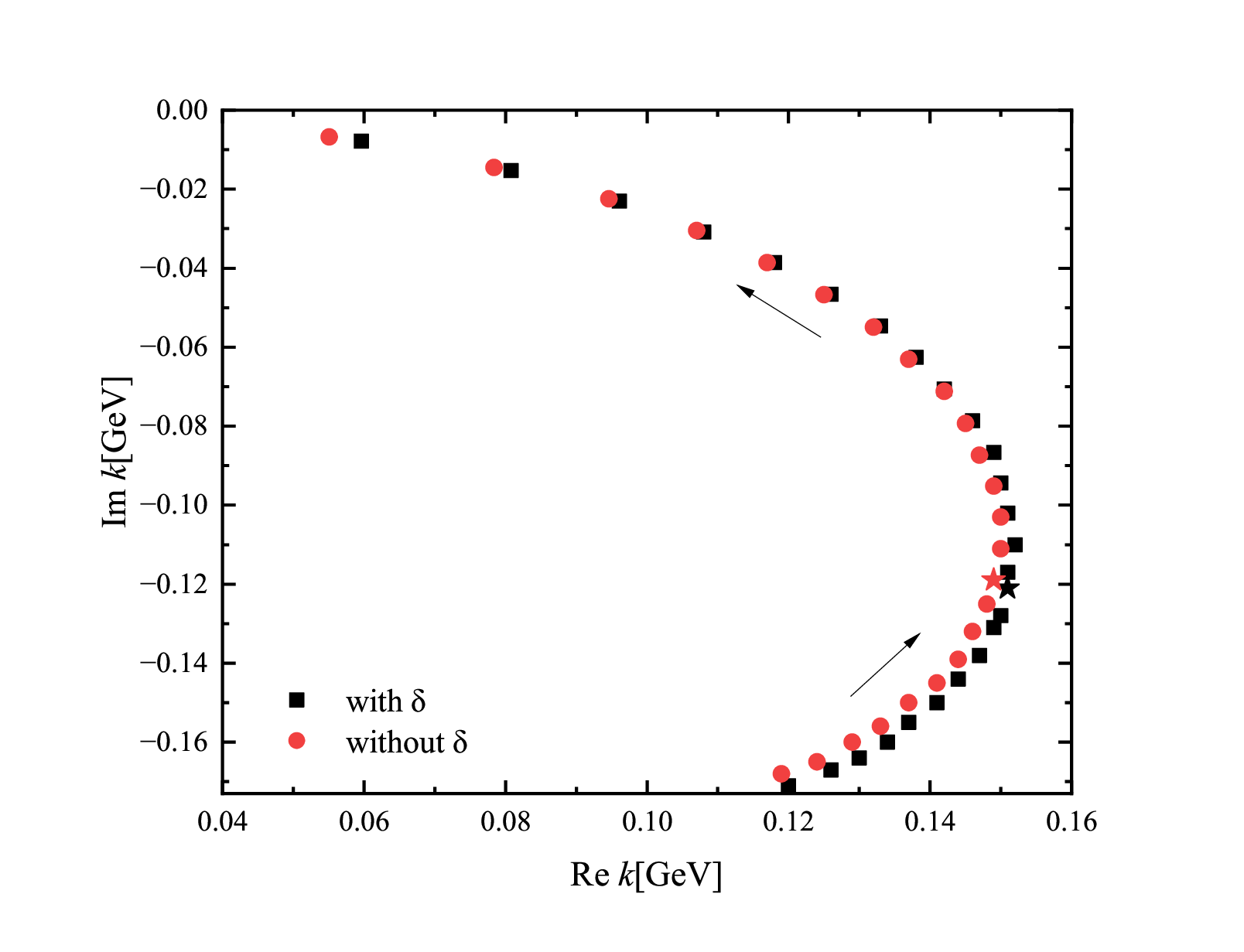}
	\caption{(Color online) The pole trajectories of $G(3900)$ with the cutoff parameters $\Lambda$ in the complex momentum plane. Black squares represent results with the contact term $\delta$, while red circles represent results without the contact term $\delta$.}
	\label{Gtrajectory}
\end{figure}	
	
In Ref.~\cite{Lin:2024prl}, the $G(3900)$ is regarded as a $P$-wave resonance, which is consistent with the quantum numbers of the $0^-(1^{--})$ state. The energy eigenvalue of a resonant state is a complex value $E=E_r-i\Gamma/2$ and appears as an isolated pole in the fourth quadrant of the complex momentum plane. The real part of the complex eigenvalue corresponds to the resonance energy, while the imaginary part equals half of the decay width. In Fig.\ref{Gtrajectory}, we present the pole trajectories of $G(3900)$ in the complex momentum plane, where the black squares and red circles represent the results with and without the contact term, respectively. As illustrated in Fig.\ref{Gtrajectory}, the pole trajectories with and without the contact term are nearly identical. With the increase of the cutoff parameter $\Lambda$, the real part of the pole energy first increases and then decreases, while the imaginary part gradually decreases, and the poles eventually merge into the continuous spectrum. Quantitatively, when the contact term is included, the pole energy evolves from (-7.73, -21.14) MeV to (5.67, -17.23) MeV, and finally approaches (1.80, -0.49) MeV. A very similar evolution is obtained without the contact term, where the pole moves from (-7.28, -20.74) MeV to (5.31, -17.14) MeV, and finally to (1.54, -0.38) MeV. The corresponding widths decrease from about 42 MeV to about 34 MeV, and eventually to below 1 MeV in both cases. This confirms that the $G(3900)$ pole gradually approaches the real energy axis and becomes a very narrow near-threshold resonance as the cutoff parameter $\Lambda$ increases.

To further elucidate the internal structure of the resonance, we combine the CMR with the Green's function method to extract the level density, scattering phase shifts, and scattering cross sections of the $P$-wave resonant state $G(3900)$. In the subsequent calculations, the cutoff parameter is fixed at $\Lambda = 0.9205$ GeV when the contact term is included, and the corresponding resonance energy is
$4.25-18.85i$ MeV. When the contact term is excluded, the cutoff parameter is adopted as $\Lambda = 1.0082$ GeV, with the resonance energy being $4.25-18.30i$ MeV. The results show that their widths are nearly identical at the same energy, and the two cutoff parameter values are also closer to each other than those of the bound states. This indicates that $P$-wave resonant states are less dependent on short-range contact terms than $S$-wave bound states. These two poles are highlighted in bold in Fig.\ref{Gtrajectory}.
	
\begin{figure}
	\centering
	\includegraphics[width=3.33in, keepaspectratio]{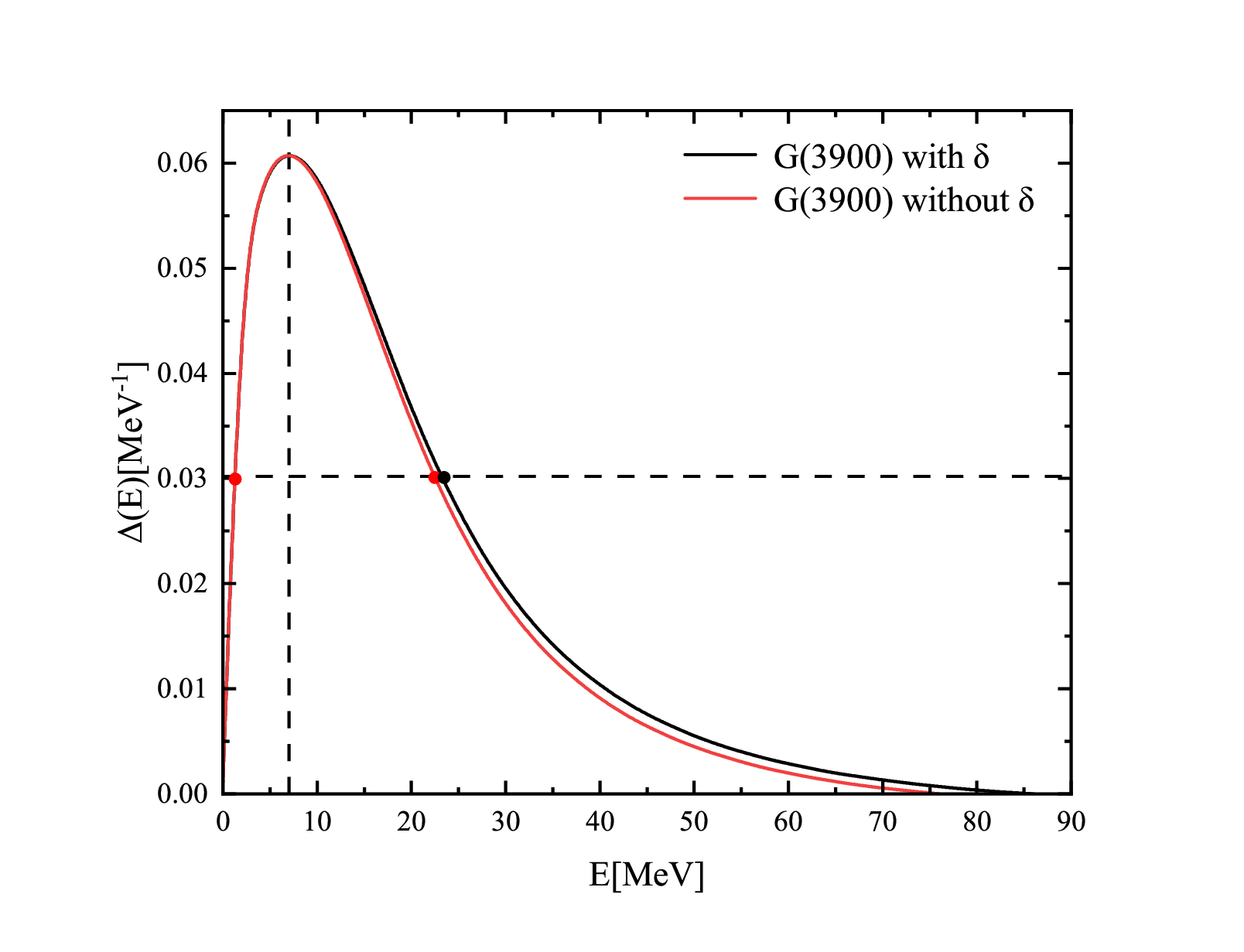}
	\caption{(Color online) The $P$-wave CLD spectrum $\Delta(E)$ of $G(3900)$ in $D\bar{D}^*/\bar{D}D^*$ scattering system, where the black solid line represents the result with the contact term $\delta$, and the red solid line represents the result without the contact term $\delta$.}
	\label{GCLD}
\end{figure}

Based on the above cutoff parameters, the $P$-wave CLD $\Delta(E)$ of G(3900) is derived from Eq.\ref{CMR-CLD} and plotted in Fig.\ref{GCLD}, in which the black and red solid lines denote the results with and without the contact term $\delta$, respectively. As shown in Fig.\ref{GCLD}, a resonance peak emerges in the CLD after background subtraction. The projection of the resonance peak onto the horizontal axis corresponds to the real energy part of the resonance state, and the distance between the two labeled intersection points is the half-height width $\Gamma$ of the CLD.
As shown in Fig.\ref{GCLD}, both the CLD distributions exhibit resonance peaks near the threshold, reach their maximum at $E \approx 6$ MeV and then gradually decrease. Meanwhile, the peak positions remain unchanged with or without the contact term $\delta$, the corresponding spectral width is slightly broader when the contact term is included. These features are consistent with the properties of the energy eigenvalues $4.25-18.85i$ and $4.25-18.30i$.

\begin{figure}
	\centering
	\includegraphics[width=3.33in, keepaspectratio]{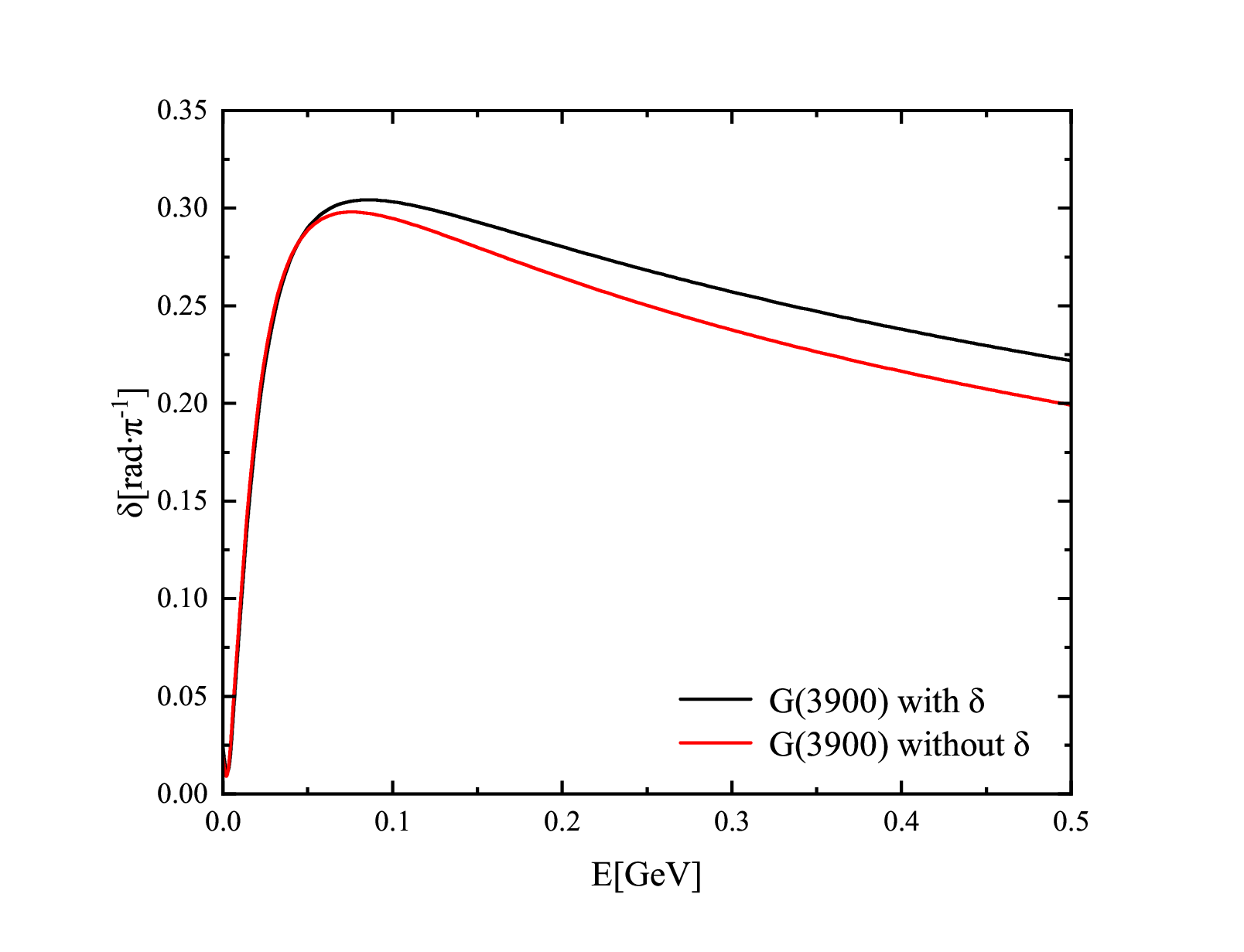}
	\caption{(Color online) Scattering shifts of $G(3900)$, where the black solid line represents the case with the contact term $\delta$ , and the red solid line represents the case without the contact term $\delta$ .}
	\label{Gshift}
\end{figure}

Scattering phase shift is an important physical quantity describing the interaction between particles. The CLD derived from the CMR Green's function method enables us to explore the $P$-wave resonance scattering phase shifts. The $P$-wave scattering phase shift $\delta(E)$ of $G(3900)$ is plotted as a function of energy $E$ in Fig.\ref{Gshift}, where the black solid line and the red solid line correspond to the results with and without the contact term $\delta$, respectively. From Fig.\ref{Gshift}, it can be observed that the scattering phase shifts with and without the short-range contact term both rise rapidly to their peaks and then decrease gradually. In the low-energy regime, the two curves are nearly coincident. The scattering phase shift including the contact term has a higher peak and decays more slowly in the high-energy region.

\begin{figure}
	\centering
	\includegraphics[width=3.33in, keepaspectratio]{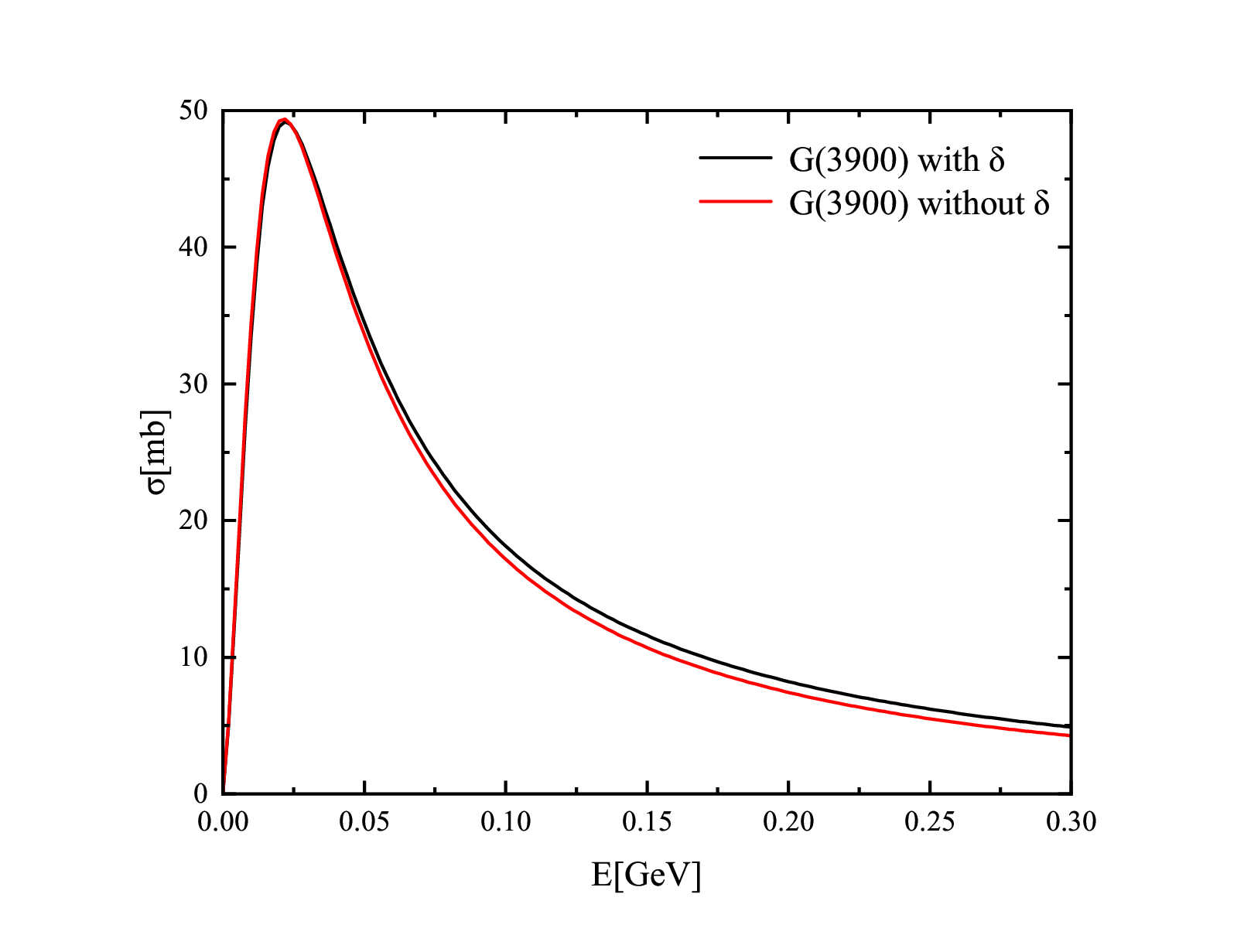}
	\caption{(Color online) Scattering cross-section of $G(3900)$, where the black solid line represents the case with the contact term $\delta$, while the red solid line represents the case without the contact term $\delta$.}
	\label{Gcross}
\end{figure}
	
From Eq.\ref{cross-section}, the scattering cross section of this process can be further derived. The scattering cross section $\sigma(E)$ of $G(3900)$ as a function of energy $E$ is presented in Fig.\ref{Gcross}, where the black and red solid lines denote the results with and without the short-range contact term $\delta$, respectively. As shown in Fig.\ref{Gcross}, the $P$-wave scattering cross section of $G(3900)$ exhibits an obvious resonance peak structure. The resonance peaks with and without the short-range contact term both appear near $0.02$ GeV. At high energies, the scattering cross section with the contact term is slightly larger, in accordance with the behaviors of CLD and scattering phase shifts.

\section{Summary}\label{sec5}
The hadronic molecular scenario provides a natural and physically intuitive framework for understanding exotic hadrons near the two-hadron thresholds. In this work, we introduce the complex momentum representation combined with the Green's function method to hadronic systems, and provide a unified and consistent description of bound states, resonant states, and the continuum spectrum.
We propose a projection operator method, which enables us to efficiently decompose and extract partial-wave components for systems with various quantum numbers and solve the Schr\"{o}dinger equation directly in momentum space. In this framework, $X(3872)$, $T_{cc}^+$, and $Z_c(3900)$ can be self-consistently interpreted as $S-D$ coupled $D\bar{D}^\ast$ bound states. Meanwhile, the $G(3900)$ state is clearly identified as a $P$-wave resonant state. By combining the CMR with the Green's function method, we further calculate several crucial physical observables, including the continuum level density, scattering phase shifts, and scattering cross sections. After background subtraction, the continuum level density shows prominent resonance peaks, whose positions and widths correspond directly to the resonance energy and decay width. The phase shifts and cross sections exhibit typical resonant behavior. Compared with $S$-wave bound states, $P$-wave resonant states are less sensitive to short-range contact interactions.

\section*{ACKNOWLEDGMENTS}
This work was supported in part by the National Natural Science Foundation of China (No.12475115), the Natural Science Foundation of Anhui Province (No.2208085MA10), and the Key Research Foundation of Education Ministry of Anhui Province of China (No.KJ2021A0061).

\appendix
\section{Hamiltonian Matrix}\label{appendix;H}
The momentum-space potential in Eq.\ref{potential} can be formulated following the approach in Ref.~\cite{Zhu:2024hgm},
\begin{equation}
V = C_{\text{coupling}} \times {\cal O}_{r,s} \times \mathcal{O}_{\text{iso}}, \nonumber
\end{equation}
where the expression of each term is presented in Table~\ref{tab:potential} and the isospin factor is shown in Table~\ref{tab:iso_factor}. The meson exchange momentum is $\vec{q} = \vec{k}' - \vec{k}$ in the direct diagram and $\vec{Q} = \vec{k}' + \vec{k}$ in the cross diagram. Here, $u$ denotes the effective mass of the exchanged meson. In the static limit, we take the effective mass to be the physical meson mass in the following. The interaction term in the cross diagram can be expanded as follows,
\[
\begin{aligned}
	& -\frac{(\vec{Q} \cdot \vec{\epsilon}'^{*})(\vec{Q} \cdot \vec{\epsilon})}{\vec{Q}^2 +\mu^2} =\frac{1}{3}(\vec{\epsilon}'^{*}\cdot\vec{\epsilon}) \cdot \left( \frac{\mu^2}{\vec{Q}^2 + \mu^2} - 1 \right) \nonumber\\
	& + \frac{1}{3}\Big( 3(\vec{\epsilon}'^{*} \cdot \hat{Q})(\vec{\epsilon} \cdot \hat{Q}) - \vec{\epsilon}'^{*}\cdot\vec{\epsilon} \Big)
	\left( -\frac{\vec{Q}^2}{\vec{Q}^2 + \mu^2} \right) \\[12pt]
\end{aligned}
\]
After expansion, the interaction can be decomposed into three terms. The momentum-space interaction is further transformed into coordinate space via Fourier transform, yielding the central potential $C(r;m)$, tensor potential $T(r;m)$, and short-range contact potential $\delta(r;m)$, which are expressed as follows:
\begin{eqnarray}
 \frac{1}{3} [ -\vec{\epsilon}'^{*}\cdot\vec{\epsilon} \delta(r;\mu)+\vec{\epsilon}'^{*}\cdot\vec{\epsilon}  C(r;\mu)
 +S_{\varepsilon_{1}^{\ast},\varepsilon_{2}} \, T(r;\mu) ].
\end{eqnarray}

Here, the tensor operators are defined as,
\begin{eqnarray}
S_{\varepsilon^{\ast},\varepsilon} &=& 3 ( \vec{\epsilon}'^{*} \!\cdot\!\hat{r} ) ( \vec{\epsilon} \!\cdot\!\hat{r} ) -  \vec{\epsilon}'^{*}\cdot\vec{\epsilon}, \nonumber
\end{eqnarray}
where, $\hat{r}=\vec{r}/r$ is the unit vector between the two mesons.
\begin{eqnarray}
\hspace{-3em}&&C(r;m) \!=\! \int \frac{\mbox{d}^{3}\vec{q}}{(2\pi)^3} \frac{m^{2}}{\vec{q}^{\,\,2}+m^{2}}
 e^{i\vec{q} \cdot \vec{r}} , \\
\hspace{-3em}&&T(r;m) S_{12}(\hat{r}) \!=\! \int \frac{\mbox{d}^{3}\vec{q}}{(2\pi)^3} \frac{- \vec{q}^{\,\,2}}{\vec{q}^{\,\,2}+m^{2}}
S_{12}(\hat{q})e^{i\vec{q} \cdot \vec{r}} ,
\end{eqnarray}
with $S_{12}(\hat{r}) \!=\! 3 (\vec{\epsilon}'^{*}\!\cdot\! \hat{r})
(\vec{\epsilon}\!\cdot\! \hat{r}) - \vec{\epsilon}'^{*} \!\cdot\!
\vec{\epsilon}$.

\begin{table}[hpt]
    \centering
      \caption{ Momentum space potentials for the \(\DD\), and \(\DDbar\) systems. The superscripts \(D\) and \(C\) denote the direct and cross diagrams, respectively. The terms \(\vec{\epsilon}'^* \) and \(\vec{\epsilon}\) represent the polarization vectors of the final and initial vector mesons, respectively. In the direct diagram, \(\vec{q} = \vec{k}' - \vec{k}\), while in the cross diagram, \(\vec{Q} = \vec{k}' + \vec{k}\), where \(\vec{k}'\) and \(\vec{k}\) are the momenta of the final and initial states, respectively. Here, \(u\) denotes the effective mass of the exchanged meson. Two additional isospin operators are defined as \(\bm{\tau}_{1} \cdot \bm{\tau}_{2}^{C} \equiv (3I - \bm{\tau}_{1} \cdot \bm{\tau}_{2})/2\) and \(I^{C} \equiv (I + \bm{\tau}_{1} \cdot \bm{\tau}_{2})/2\).  }
    \label{tab:potential}
 \begin{tabular*}{\hsize}{@{}@{\extracolsep{\fill}}cccccccccc@{}}
\hline
\hline
\multirow{2}{*}{$V$}& \multirow{2}{*}{$C_{\text{coupling}}$} & \multirow{2}{*}{$\mathcal{O}_{r,s}$} & \multicolumn{3}{c}{$\mathcal{O}_{iso}$}\tabularnewline
 & & &  $DD^{*}$ & $[\DDbar]^{C=+1}$ & $[\DDbar]^{C=-1}$\tabularnewline
\hline
$V_{\rho}^{D}$ & $\frac{\beta^{2}g_{v}^{2}}{2}$ &  \multirow{3}{*}{$-\frac{\vec{\epsilon}'^{*}\cdot\vec{\epsilon}}{\vec{q}^{2}+u^{2}}$} & $-\frac{\bm{\tau}_{1}\cdot\boldsymbol{\tau}_{2}}{2}$  & $-\frac{\bm{\tau}_{1}\cdot\bm{\tau}_{2}}{2}$ & $-\frac{\bm{\tau}_{1}\cdot\bm{\tau}_{2}}{2}$\tabularnewline
$V_{\omega}^{D}$ & $\frac{\beta^{2}g_{v}^{2}}{2}$ &  & $-\frac{1}{2}I$ & $\frac{1}{2}I$ & $\frac{1}{2}I$\tabularnewline
$V_{\sigma}^{D}$ & $g_{s}^{2}$ &  & $I$ & $I$ & $I$\tabularnewline
\hline
$V_{\pi}^{C}$ & $\frac{g_a^{2}}{f_{\pi}^{2}}$ &  \multirow{2}{*}{$\frac{\left(\vec{Q}\cdot\vec{\epsilon}'^{*}\right)\left(\vec{Q}\cdot\epsilon\right)}{\vec{Q}^{2}+u^{2}}$} & $-\frac{\bm{\tau}_{1}\cdot\bm{\tau}_{2}^{C}}{2}$ &  $\frac{\bm{\tau}_{1}\cdot\bm{\tau}_{2}}{2}$ & $-\frac{\bm{\tau}_{1}\cdot\bm{\tau}_{2}}{2}$\tabularnewline
$V_{\eta}^{C}$ & $\frac{g_a^{2}}{f_{\pi}^{2}}$ &  & $-\frac{1}{6}I^{C}$ & $-\frac{1}{6}I$ & $\frac{1}{6}I$\tabularnewline
$V_{\rho}^{C}$ & $2\lambda^{2}g_{v}^{2}$ & \multirow{2}{*}{$\frac{\left(\vec{Q}\cdot\bm{\epsilon}'^{*}\right)\left(\vec{Q}\cdot\epsilon\right)-\vec{Q}^{2}\left(\vec{\epsilon}'^{*}\cdot\vec{\epsilon}\right)}{\vec{Q}^{2}+u^{2}}$} & $\frac{\bm{\tau}_{1}\cdot\bm{\tau}_{2}^{C}}{2}$ & $\frac{\bm{\tau}_{1}\cdot\bm{\tau}_{2}}{2}$ & $-\frac{\bm{\tau}_{1}\cdot\bm{\tau}_{2}}{2}$\tabularnewline
$V_{\omega}^{C}$ & $2\lambda^{2}g_{v}^{2}$ &  & $\frac{1}{2}I^{C}$ & $-\frac{1}{2}I$ & $\frac{1}{2}I$\tabularnewline
\hline
\hline
\end{tabular*}
\end{table}

\begin{table}[]
    \centering
        \caption{Isospin factor for $\DDbar$ and $\DD$ systems. }
    \label{tab:iso_factor}
\begin{tabular*}{\hsize}{@{}@{\extracolsep{\fill}}ccccccccccccc@{}}
\hline
\hline
 & \multicolumn{2}{c}{$DD^{*}$} & \multicolumn{2}{c}{$\DDbar,C=+1$} & \multicolumn{2}{c}{$\DDbar,C=-1$}\tabularnewline
$\langle\mathcal{O}_{iso}\rangle$ & $I=0$ & $I=1$ & $I=0$ & $I=1$ & $I=0$ & $I=1$\tabularnewline
\hline
$V_{\rho}^{D}$ & $\frac{3}{2}$ & $-\frac{1}{2}$ & $\frac{3}{2}$ & $-\frac{1}{2}$ & $\frac{3}{2}$ & $-\frac{1}{2}$\tabularnewline
$V_{\omega}^{D}$ & $-\frac{1}{2}$ & $-\frac{1}{2}$ & $\frac{1}{2}$ & $\frac{1}{2}$  & $\frac{1}{2}$ & $\frac{1}{2}$\tabularnewline
$V_{\sigma}^{D}$ & $1$ & $1$ & $1$ & $1$ & $1$ & $1$\tabularnewline
\hline
$V_{\pi}^{C}$ & $-\frac{3}{2}$ & $-\frac{1}{2}$ & $-\frac{3}{2}$ & $\frac{1}{2}$ & $\frac{3}{2}$ & $-\frac{1}{2}$\tabularnewline
$V_{\eta}^{C}$ & $\frac{1}{6}$ & $-\frac{1}{6}$ & $-\frac{1}{6}$ & $-\frac{1}{6}$ & $\frac{1}{6}$ & $\frac{1}{6}$\tabularnewline
$V_{\rho}^{C}$ & $\frac{3}{2}$ & $\frac{1}{2}$ & $-\frac{3}{2}$ & $\frac{1}{2}$ & $\frac{3}{2}$ & $-\frac{1}{2}$\tabularnewline
$V_{\omega}^{C}$ & $-\frac{1}{2}$ & $\frac{1}{2}$ & $-\frac{1}{2}$ & $-\frac{1}{2}$ & $\frac{1}{2}$ & $\frac{1}{2}$\tabularnewline
\hline
\hline
\end{tabular*}
\end{table}

In the following, we define $|\vec{k}|=k$, ~$|\vec{k}'|=k'$,~$\vec{n}=\vec{k}/k$,~$\vec{n}'=\vec{k}'/k'$, and $x=\vec{n}\cdot\vec{n}'=cos\theta$. Thus, $\vec{q}^2 = k'^2 + k^2-2kk'x$ and $\vec{Q}^2 = k'^2 + k^2+2kk'x$. By employing the projection operator method presented in Eq.~\ref{project}, the momentum-space potentials for various quantum numbers can be obtained as follows. It should be noted that the final results are the following expressions with form factors included.

\begin{align*}
V_{G(3900)}^{with} &= \langle^3P_1|V|^3P_1\rangle = \\
    & \frac{1}{2} \beta^2 g_V^2 \int_{-1}^{1}\frac{dx}{2} \left( -\frac{x}{\vec{q}^2 + m_\rho^2} \right) \left(-\frac{\bm{\tau_1} \cdot \bm{\tau_2}}{2}\right) \\
	&+ \frac{1}{2} \beta^2 g_V^2 \int_{-1}^{1}\frac{dx}{2} \left( -\frac{x}{\vec{q}^{2} + m_\omega^2} \right) \left( -\frac{1}{2} \bm{I} \right) \\
	&+ g_s^2 \int_{-1}^{1}\frac{dx}{2} \left( -\frac{x}{\vec{q}^{2} + m_\sigma^2} \right) \\
	&+ \frac{g_a^2}{f_\pi^2} \int_{-1}^{1}\frac{dx}{2} \frac{\frac{1}{2}\left(kk'x^2 - kk'\right)}{\vec{Q}^2 + m_\pi^2} \, \left(-\frac{\bm{\tau_1} \cdot \bm{\tau_2}}{2}\right) \\
	&+ \frac{g_a^2}{f_\pi^2} \int_{-1}^{1}\frac{dx}{2} \frac{\frac{1}{2}\left(kk'x^2 - kk'\right)}{\vec{Q}^2 + m_\eta^2}\left(-\frac{1}{6} \bm{I}^C\right) \\
	&+ 2\lambda^2 g_V^2 \int_{-1}^{1}\frac{dx}{2}  \frac{\frac{1}{2}\left(kk'x^2 - kk'\right)-\vec{Q}^2x}{\vec{Q}^2 + m_\rho^2} \, \left(\frac{\bm{\tau_1} \cdot \bm{\tau_2}}{2}\right) \\
	&+ 2\lambda^2 g_V^2 \int_{-1}^{1}\frac{dx}{2}  \frac{\frac{1}{2}\left(kk'x^2 - kk'\right)-\vec{Q}^2x}{\vec{Q}^2 + m_\omega^2} \,\left(\frac{1}{2}\bm{I}^C\right)
\end{align*}
\[
\begin{aligned}
	&V_{G(3900)}^{without} =\langle^3P_1|V|^3P_1\rangle \\
	&\quad= \frac{1}{2}\beta^2 g_V^2 \int_{-1}^{1} \frac{dx}{2}
	\left( -\frac{x}{\vec{q}^2 + m_\rho^2} \right)
	\left(-\frac{\bm{\tau_1} \cdot \bm{\tau_2}}{2}\right) \\[6pt]
	&\quad + \frac{1}{2}\beta^2 g_V^2 \int_{-1}^{1} \frac{dx}{2}
	\left( -\frac{x}{\vec{q}^2 + m_\omega^2} \right)
	\left( -\frac{1}{2}\bm{I} \right)
	+ g_s^2 \int_{-1}^{1} \frac{dx}{2} \frac{-x}{\vec{q}^2 + m_\sigma^2} \\[6pt]
	&\quad + \frac{g_a^2}{f_\pi^2} \int_{-1}^{1} \frac{dx}{2}
	\left( -\frac{1}{3}x \right)
	\left(-\frac{\bm{\tau_1} \cdot \bm{\tau_2}}{2}\right) \\[6pt]
	&\quad + \frac{g_a^2}{f_\pi^2} \int_{-1}^{1} \frac{dx}{2}
	\frac{\frac{1}{2}(kk'x^2-kk')}{\vec{Q}^2 + m_\pi^2}
	\left(-\frac{\bm{\tau_1} \cdot \bm{\tau_2}}{2}\right) \\[6pt]
	&\quad + \frac{g_a^2}{f_\pi^2} \int_{-1}^{1} \frac{dx}{2}
	\left( -\frac{1}{3}x \right)
	\left( -\frac{1}{6}\bm{I}^C \right) \\[6pt]
	&\quad + \frac{g_a^2}{f_\pi^2} \int_{-1}^{1} \frac{dx}{2}
	\frac{\frac{1}{2}(kk'x^2-kk')}{\vec{Q}^2 + m_\eta^2}
	\left( -\frac{1}{6}\bm{I}^C \right) \\[6pt]
	&\quad + 2\lambda^2 g_V^2 \cdot \int_{-1}^{1} \frac{dx}{2}
	\left( -\frac{1}{3}x \right)
	\left(\frac{\bm{\tau_1} \cdot \bm{\tau_2}}{2}\right) \\[6pt]
	&\quad + 2\lambda^2 g_V^2 \int_{-1}^{1} \frac{dx}{2}
	\frac{\frac{1}{2}(kk'x^2-kk')}{\vec{Q}^2 + m_\rho^2}
	\left(\frac{\bm{\tau_1} \cdot \bm{\tau_2}}{2}\right) \\[6pt]
	&\quad + 2\lambda^2 g_V^2 \cdot \int_{-1}^{1} \frac{dx}{2}
	\frac{xm_\rho^2}{\vec{Q}^2 + m_\rho^2}
	\left(\frac{\bm{\tau_1} \cdot \bm{\tau_2}}{2}\right) \\[6pt]
	&\quad + 2\lambda^2 g_V^2 \cdot \int_{-1}^{1} \frac{dx}{2}
	\left( -\frac{1}{3}x \right)
	\left( \frac{1}{2}\bm{I}^C \right) \\[6pt]
	&\quad + 2\lambda^2 g_V^2 \cdot \int_{-1}^{1} \frac{dx}{2}
	\frac{\frac{1}{2}(kk'x^2-kk')}{\vec{Q}^2 + m_\omega^2}
	\left( \frac{1}{2}\bm{I}^C \right) \\[6pt]
	&\quad + 2\lambda^2 g_V^2 \cdot \int_{-1}^{1} \frac{dx}{2}
	\frac{xm_\omega^2}{\vec{Q}^2 + m_\omega^2}
	\left( \frac{1}{2}\bm{I}^C \right)
\end{aligned}
\]

\[
V_{X(3872)/T_{cc}^+/Z(3900)}^{with} = \begin{pmatrix}
	\langle^3S_1|V^{with}|^3S_1\rangle & \langle^3S_1|V^{with}|^3D_1\rangle  \\
	\langle^3D_1|V^{with}|^3S_1\rangle  & \langle^3D_1|V^{with}|^3D_1\rangle
\end{pmatrix}
\]
\[
\begin{aligned}
	&\langle^3S_1|V^{with}|^3S_1\rangle = \\[6pt]
	&\quad \frac{1}{2}\beta^2 g_v^2 \int_{-1}^{1} \frac{dx}{2}
	\left( -\frac{1}{\vec{q}^2 + m_\rho^2} \right)
	\cdot \left(-\frac{\bm{\tau_1} \cdot \bm{\tau_2}}{2}\right) \\[6pt]
	&\quad + \frac{1}{2}\beta^2 g_v^2 \int_{-1}^{1} \frac{dx}{2}
	\left( -\frac{1}{\vec{q}^2 + m_\omega^2} \right)
	\cdot \left( -\frac{1}{2}\bm{I} \right)\\[6pt]
	&\quad + g_s^2 \int_{-1}^{1} \frac{dx}{2}
	\left( -\frac{1}{\vec{q}^2 + m_\sigma^2} \right) \\[6pt]
	&\quad + \frac{g_a^2}{f_\pi^2} \int_{-1}^{1} \frac{dx}{2}
	\frac{\frac{1}{3}\vec{Q}^2}{\vec{Q}^2 + m_\pi^2}
	\cdot \left(-\frac{\bm{\tau_1} \cdot \bm{\tau_2}}{2}\right) \\[6pt]
	&\quad + \frac{g_a^2}{f_\pi^2} \int_{-1}^{1} \frac{dx}{2}
	\frac{\frac{1}{3}\vec{Q}^2}{\vec{Q}^2 + m_\eta^2}
	\cdot \left( -\frac{1}{6}\bm{I}^C \right) \\[6pt]
	&\quad + 2\lambda^2 g_v^2 \int_{-1}^{1} \frac{dx}{2}
	\frac{-\frac{2}{3}\vec{Q}^2}{\vec{Q}^2 + m_\rho^2}
	\cdot \left(\frac{\bm{\tau_1} \cdot \bm{\tau_2}}{2}\right) \\[6pt]
	&\quad + 2\lambda^2 g_v^2 \int_{-1}^{1} \frac{dx}{2}
	\frac{-\frac{2}{3}\vec{Q}^2 }{\vec{Q}^2 + m_\omega^2}
	\cdot \left( \frac{1}{2}\bm{I}^C \right)
\end{aligned}
\]

\[
\begin{aligned}
	&\langle^3S_1|V^{with}|^3D_1\rangle = \\[6pt]
	&\quad\frac{g_a^2}{f_\pi^2} \int_{-1}^{1} \frac{dx}{2}
	\left( -\frac{1}{\sqrt{18}} \right)
	\frac{3(k' + kx)^2 - \vec{Q}^2}{\vec{Q}^2 + m_\pi^2}
	\left(-\frac{\bm{\tau_1} \cdot \bm{\tau_2}}{2}\right) \\[6pt]
	&\quad + \frac{g_a^2}{f_\pi^2} \int_{-1}^{1} \frac{dx}{2}
	\left( -\frac{1}{\sqrt{18}} \right)
	\frac{3(k' + kx)^2 - \vec{Q}^2}{\vec{Q}^2 + m_\eta^2}
	\left( -\frac{1}{6}\bm{I}^C \right) \\[6pt]
	&\quad + 2\lambda^2 g_v^2 \int_{-1}^{1} \frac{dx}{2}
	\left( -\frac{1}{\sqrt{18}} \right)
	\frac{3(k' + kx)^2 - \vec{Q}^2}{\vec{Q}^2 + m_\rho^2}
	\left(\frac{\bm{\tau_1} \cdot \bm{\tau_2}}{2}\right) \\[6pt]
	&\quad + 2\lambda^2 g_v^2 \int_{-1}^{1} \frac{dx}{2}
	\left( -\frac{1}{\sqrt{18}} \right)
	\frac{3(k' + kx)^2 - \vec{Q}^2}{\vec{Q}^2 + m_\omega^2}
	\left( \frac{1}{2}\bm{I}^C \right)
\end{aligned}
\]

\[
\begin{aligned}
	& \langle^3D_1|V^{with}|^3S_1\rangle = \\[6pt]
	&\frac{g_a^2}{f_\pi^2} \int_{-1}^{1} \frac{dx}{2}
	\left( -\frac{1}{\sqrt{18}} \right)
	\frac{3(k + k'x)^2 - \vec{Q}^2}{\vec{Q}^2 + m_\pi^2}
	\left(-\frac{\bm{\tau_1} \cdot \bm{\tau_2}}{2}\right) \\[6pt]
	&\quad + \frac{g_a^2}{f_\pi^2} \int_{-1}^{1} \frac{dx}{2}
	\left( -\frac{1}{\sqrt{18}} \right)
	\frac{3(k + k'x)^2 - \vec{Q}^2}{\vec{Q}^2 + m_\eta^2}
	\left( -\frac{1}{6}\bm{I}^C \right) \\[6pt]
	&\quad + 2\lambda^2 g_v^2 \int_{-1}^{1} \frac{dx}{2}
	\left( -\frac{1}{\sqrt{18}} \right)
	\frac{3(k + k'x)^2 - \vec{Q}^2}{\vec{Q}^2 + m_\rho^2}
	\left(\frac{\bm{\tau_1} \cdot \bm{\tau_2}}{2}\right) \\[6pt]
	&\quad + 2\lambda^2 g_v^2 \int_{-1}^{1} \frac{dx}{2}
	\left( -\frac{1}{\sqrt{18}} \right)
	\frac{3(k + k'x)^2 - \vec{Q}^2}{\vec{Q}^2 + m_\omega^2}
	\left( \frac{1}{2}\bm{I}^C \right)
\end{aligned}
\]

\[
\begin{aligned}
	&\langle^3D_1|V^{with}|^3D_1\rangle = \\[6pt]
	&\quad \frac{1}{2}\beta^2 g_v^2
	\int_{-1}^{1} \frac{dx}{2} \frac{3x^2-1}{2}
	 \frac{-1}{\vec{q}^2 + m_\rho^2}
	\left(-\frac{\bm{\tau_1} \cdot \bm{\tau_2}}{2}\right) \\[6pt]
	&\quad+ \frac{1}{2}\beta^2 g_v^2
	\int_{-1}^{1} \frac{dx}{2} \frac{3x^2-1}{2}
	\cdot \frac{-1}{\vec{q}^2 + m_\omega^2}
	\left(-\frac{1}{2}\bm{I}\right) \\[6pt]
	&\quad+ g_s^2
	\int_{-1}^{1} \frac{dx}{2} \frac{3x^2-1}{2}
	 \frac{-1}{\vec{q}^2 + m_\sigma^2} \\[6pt]
	&\quad+ \frac{g_a^2}{f_\pi^2} \left(-\frac{1}{6}\right)
	\left(-\frac{\bm{\tau_1} \cdot \bm{\tau_2}}{2}\right) \\[6pt]
	&\quad\times \int_{-1}^{1} \frac{dx}{2}
	\frac{2k^2+2k'^2+kk'x-6k^2x^2-6k'^2x^2-9kk'x^3}
	{\vec{Q}^2 + m_\pi^2} \\[6pt]
	&\quad+ \frac{g_a^2}{f_\pi^2} \left(-\frac{1}{6}\right)
	\left(-\frac{1}{6}\bm{I}^C\right) \\[6pt]
	&\quad\times \int_{-1}^{1} \frac{dx}{2}
	\frac{2k^2+2k'^2+kk'x-6k^2x^2-6k'^2x^2-9kk'x^3}
	{\vec{Q}^2 + m_\eta^2} \\[6pt]
	&\quad+ 2\lambda^2 g_v^2 \left(-\frac{1}{6}\right)
	\left(\frac{\bm{\tau_1} \cdot \bm{\tau_2}}{2}\right) \\[6pt]
	&\quad\times \int_{-1}^{1} \frac{dx}{2}
	\frac{2k^2+2k'^2+kk'x-6k^2x^2-6k'^2x^2-9kk'x^3}
	{\vec{Q}^2 + m_\rho^2} \\[6pt]
	&\quad+ 2\lambda^2 g_v^2
	\int_{-1}^{1} \frac{dx}{2} \frac{3x^2-1}{2}
	\frac{-\vec{Q}^2}{\vec{Q}^2 + m_\rho^2}
	\left(\frac{\bm{\tau_1} \cdot \bm{\tau_2}}{2}\right) \\[6pt]
	&+ 2\lambda^2 g_v^2 \left(-\frac{1}{6}\right)
	\left(\frac{1}{2}\bm{I}^C\right) \\[6pt]
	&\quad\times \int_{-1}^{1} \frac{dx}{2}
	\frac{2k^2+2k'^2+kk'x-6k^2x^2-6k'^2x^2-9kk'x^3}
	{\vec{Q}^2 + m_\omega^2} \\[6pt]
	&\quad+ 2\lambda^2 g_v^2
	\int_{-1}^{1} \frac{dx}{2} \frac{3x^2-1}{2}
	\frac{-\vec{Q}^2}{\vec{Q}^2 + m_\omega^2}
	\left(\frac{1}{2}\bm{I}^C\right)
\end{aligned}
\]

\[
V_{X(3872)/T_{cc}^+/Z(3900)}^{with} = \begin{pmatrix}
	\langle^3S_1|V^{without}|^3S_1\rangle & \langle^3S_1|V^{without}|^3D_1\rangle  \\
	\langle^3D_1|V^{without}|^3S_1\rangle  & \langle^3D_1|V^{without}|^3D_1\rangle
\end{pmatrix}
\]

\[
\begin{aligned}
	&\langle^3S_1|V^{without}|^3S_1\rangle = \\[6pt]
	&\quad\frac{1}{2}\beta^2 g_v^2
	\int_{-1}^{1} \frac{dx}{2}
	\left( -\frac{1}{\vec{q}^2 + m_\rho^2} \right)
	\left(-\frac{\bm{\tau_1} \cdot \bm{\tau_2}}{2}\right) \\[6pt]
	&\quad + \frac{1}{2}\beta^2 g_v^2
	\int_{-1}^{1} \frac{dx}{2}
	\left( -\frac{1}{\vec{q}^2 + m_\omega^2} \right)
	\left( -\frac{1}{2}\bm{I} \right) \\[6pt]
	&\quad + g_s^2 \int_{-1}^{1} \frac{dx}{2}
	\left( -\frac{1}{\vec{q}^2 + m_\sigma^2} \right) \\[6pt]
	&\quad + \frac{g_a^2}{f_\pi^2}
	\int_{-1}^{1} \frac{dx}{2}
	\left( -\frac{\frac{1}{3}m_\pi^2}{\vec{Q}^2 + m_\pi^2} \right)
	\left(-\frac{\bm{\tau_1} \cdot \bm{\tau_2}}{2}\right) \\[6pt]
	&\quad + \frac{g_a^2}{f_\pi^2}
	\int_{-1}^{1} \frac{dx}{2}
	\left( -\frac{\frac{1}{3}m_\eta^2}{\vec{Q}^2 + m_\eta^2} \right)
	\left( -\frac{1}{6}\bm{I}^C \right) \\[6pt]
	&\quad + 2\lambda^2 g_v^2
	\int_{-1}^{1} \frac{dx}{2}
	\left( \frac{\frac{2}{3}m_\rho^2}{\vec{Q}^2 + m_\rho^2} \right)
	\left(\frac{\bm{\tau_1} \cdot \bm{\tau_2}}{2}\right) \\[6pt]
	&\quad + 2\lambda^2 g_v^2
	\int_{-1}^{1} \frac{dx}{2}
	\left( \frac{\frac{2}{3}m_\omega^2}{\vec{Q}^2 + m_\omega^2} \right)
	\left( \frac{1}{2}\bm{I}^C \right)
\end{aligned}
\]

\[
\begin{aligned}
	&\langle^3S_1|V^{without}|^3D_1\rangle = \\[6pt]
	&\quad \frac{g_a^2}{f_\pi^2}
	\int_{-1}^{1} \frac{dx}{2}
	\left( -\frac{1}{\sqrt{18}} \right)
	\frac{3(k' + kx)^2 - \vec{Q}^2}{\vec{Q}^2 + m_\pi^2}
	\left(-\frac{\bm{\tau_1} \cdot \bm{\tau_2}}{2}\right) \\[6pt]
	&\quad + \frac{g_a^2}{f_\pi^2}
	\int_{-1}^{1} \frac{dx}{2}
	\left( -\frac{1}{\sqrt{18}} \right)
	\frac{3(k' + kx)^2 - \vec{Q}^2}{\vec{Q}^2 + m_\eta^2}
	\left( -\frac{1}{6}\bm{I}^C \right) \\[6pt]
	&\quad + 2\lambda^2 g_v^2
	\int_{-1}^{1} \frac{dx}{2}
	\left( -\frac{1}{\sqrt{18}} \right)
	\frac{3(k' + kx)^2 - \vec{Q}^2}{\vec{Q}^2 + m_\rho^2}
	\left(\frac{\bm{\tau_1} \cdot \bm{\tau_2}}{2}\right) \\[6pt]
	&\quad + 2\lambda^2 g_v^2
	\int_{-1}^{1} \frac{dx}{2}
	\left( -\frac{1}{\sqrt{18}} \right)
	\frac{3(k' + kx)^2 - \vec{Q}^2}{\vec{Q}^2 + m_\omega^2}
	\left( \frac{1}{2}\bm{I}^C \right)
\end{aligned}
\]

\[
\begin{aligned}
	&\langle^3D_1|V^{without}|^3S_1\rangle = \\[6pt]
	&\quad \frac{g_a^2}{f_\pi^2}
	\int_{-1}^{1} \frac{dx}{2}
	\left( -\frac{1}{\sqrt{18}} \right)
	\frac{3(k + k'x)^2 - \vec{Q}^2}{\vec{Q}^2 + m_\pi^2}
	\left(-\frac{\bm{\tau_1} \cdot \bm{\tau_2}}{2}\right) \\[6pt]
	&\quad + \frac{g_a^2}{f_\pi^2}
	\int_{-1}^{1} \frac{dx}{2}
	\left( -\frac{1}{\sqrt{18}} \right)
	\frac{3(k + k'x)^2 - \vec{Q}^2}{\vec{Q}^2 + m_\eta^2}
	\left( -\frac{1}{6}\bm{I}^C \right) \\[6pt]
	&\quad + 2\lambda^2 g_v^2
	\int_{-1}^{1} \frac{dx}{2}
	\left( -\frac{1}{\sqrt{18}} \right)
	\frac{3(k + k'x)^2 - \vec{Q}^2}{\vec{Q}^2 + m_\rho^2}
	\left(\frac{\bm{\tau_1} \cdot \bm{\tau_2}}{2}\right) \\[6pt]
	&\quad + 2\lambda^2 g_v^2
	\int_{-1}^{1} \frac{dx}{2}
	\left( -\frac{1}{\sqrt{18}} \right)
	\frac{3(k + k'x)^2 - \vec{Q}^2}{\vec{Q}^2 + m_\omega^2}
	\left( \frac{1}{2}\bm{I}^C \right)
\end{aligned}
\]

\[
\begin{aligned}
	&\langle^3D_1|V^{without}|^3D_1\rangle = \\[6pt]
	&\quad \frac{1}{2}\beta^2 g_v^2
	\int_{-1}^{1} \frac{dx}{2} \frac{3x^2-1}{2}
	\left(-\frac{1}{\vec{q}^2 + m_\rho^2}\right)
	\left(-\frac{\bm{\tau_1} \cdot \bm{\tau_2}}{2}\right) \\[6pt]
	&\quad + \frac{1}{2}\beta^2 g_v^2
	\int_{-1}^{1} \frac{dx}{2} \frac{3x^2-1}{2}
	\left(-\frac{1}{\vec{q}^2 + m_\omega^2}\right)
	\left(-\frac{1}{2}\bm{I}\right) \\[6pt]
	&\quad + g_s^2
	\int_{-1}^{1} \frac{dx}{2} \frac{3x^2-1}{2}
	\left(-\frac{1}{\vec{q}^2 + m_\sigma^2}\right) \\[6pt]
	&\quad + \frac{g_a^2}{f_\pi^2}
	\int_{-1}^{1} \frac{dx}{2} \frac{3x^2-1}{2}
	\left(-\frac{1}{3}\right)
	\left(-\frac{\bm{\tau_1} \cdot \bm{\tau_2}}{2}\right) \\[6pt]
	&\quad + \frac{g_a^2}{f_\pi^2} \left(-\frac{1}{6}\right)
	\left(-\frac{\bm{\tau_1} \cdot \bm{\tau_2}}{2}\right) \\[6pt]
	&\quad \times \int_{-1}^{1} \frac{dx}{2}
	\frac{2k^2 + 2k'^2 + kk'x - 6k^2x^2 - 6k'^2x^2 - 9kk'x^3}
	{\vec{Q}^2 + m_\pi^2} \\[6pt]
	&\quad + \frac{g_a^2}{f_\pi^2}
	\int_{-1}^{1} \frac{dx}{2} \frac{-3x^2+1}{6}
	\left(-\frac{1}{6}\bm{I}^C\right) \\[6pt]
	&\quad + \frac{g_a^2}{f_\pi^2} \left(-\frac{1}{6}\right)
	\left(-\frac{1}{6}\bm{I}^C\right) \\[6pt]
	&\quad \times \int_{-1}^{1} \frac{dx}{2}
	\frac{2k^2 + 2k'^2 + kk'x - 6k^2x^2 - 6k'^2x^2 - 9kk'x^3}
	{\vec{Q}^2 + m_\eta^2} \\[6pt]
	&\quad + 2\lambda^2 g_v^2
	\int_{-1}^{1} \frac{dx}{2} \frac{-3x^2+1}{6}
	\left(\frac{\bm{\tau_1} \cdot \bm{\tau_2}}{2}\right) \\[6pt]
	&\quad + 2\lambda^2 g_v^2 \left(-\frac{1}{6}\right)
	\left(\frac{\bm{\tau_1} \cdot \bm{\tau_2}}{2}\right) \\[6pt]
	&\quad \times \int_{-1}^{1} \frac{dx}{2}
	\frac{2k^2 + 2k'^2 + kk'x - 6k^2x^2 - 6k'^2x^2 - 9kk'x^3}
	{\vec{Q}^2 + m_\rho^2} \\[6pt]
	&\quad + 2\lambda^2 g_v^2
	\int_{-1}^{1} \frac{dx}{2} \frac{3x^2-1}{2}
	\left(\frac{m_\rho^2}{\vec{Q}^2 + m_\rho^2}\right)
	\left(\frac{\bm{\tau_1} \cdot \bm{\tau_2}}{2}\right) \\[6pt]
	&\quad + 2\lambda^2 g_v^2
	\int_{-1}^{1} \frac{dx}{2} \frac{3x^2-1}{2}
	\left(-\frac{1}{3}\right)
	\left(\frac{1}{2}\bm{I}^C\right) \\[6pt]
	&\quad + 2\lambda^2 g_v^2 \left(-\frac{1}{6}\right)
	\left(\frac{1}{2}\bm{I}^C\right) \\[6pt]
	&\quad \times \int_{-1}^{1} \frac{dx}{2}
	\frac{2k^2 + 2k'^2 + kk'x - 6k^2x^2 - 6k'^2x^2 - 9kk'x^3}
	{\vec{Q}^2 + m_\omega^2} \\[6pt]
	&\quad + 2\lambda^2 g_v^2
	\int_{-1}^{1} \frac{dx}{2} \frac{3x^2-1}{2}
	\left(\frac{m_\omega^2}{\vec{Q}^2 + m_\omega^2}\right)
	\left(\frac{1}{2}\bm{I}^C\right)
\end{aligned}
\]

\end{document}